\documentclass[reqno,10pt,a4paper,dvips]{amsart}

\usepackage{amssymb,mathptmx,cite,psfrag,eucal,array,setspace,color,geometry,enumitem}
\usepackage[dvips]{graphicx}
\usepackage{tikz}

\DeclareSymbolFont{largesymbols}{OMX}{zplm}{m}{n} %Replaces summation symbol in times by the palatino one...

\numberwithin{equation}{section}

\newcolumntype{C}{>{$}c<{$}} %Defines math mode in tabular (array package)...

\allowdisplaybreaks

\newcommand{\alg}[1]{\mathfrak{#1}}
\newcommand{\grp}[1]{\mathsf{#1}}

\newcommand{\func}[2]{#1 \left( #2 \right)}
\newcommand{\tfunc}[2]{#1 \bigl( #2 \bigr)}

\newcommand{\brac}[1]{\left( #1 \right)}
\newcommand{\tbrac}[1]{\bigl( #1 \bigr)}
\newcommand{\sqbrac}[1]{\left[ #1 \right]}
\newcommand{\set}[1]{\left\{ #1 \right\}}

\newcommand{\abs}[1]{\left| #1 \right|}

\newcommand{\ZZ}{\mathbb{Z}}
\newcommand{\NN}{\mathbb{N}}
\newcommand{\RR}{\mathbb{R}}
\newcommand{\CC}{\mathbb{C}}

\newcommand{\dd}{\mathrm{d}}
\newcommand{\ii}{\mathfrak{i}}
\newcommand{\ee}{\mathsf{e}}

\newcommand{\killing}[2]{\kappa \bigl( #1 , #2 \bigr)}

\newcommand{\comm}[2]{\bigl[ #1 , #2 \bigr]}

\newcommand{\ket}[1]{\bigl\lvert #1 \bigr\rangle}
\newcommand{\braket}[2]{\bigl\langle #1 \bigr\rvert \bigl. #2 \bigr\rangle}
\newcommand{\finite}[1]{\overline{#1}}
\newcommand{\affine}[1]{\widehat{#1}}

\newcommand{\FinDiscMod}[1]{\finite{\mathcal{D}}_{#1}}
\newcommand{\FinIrrMod}[1]{\finite{\mathcal{L}}_{#1}}
\newcommand{\FinTypMod}[1]{\finite{\mathcal{E}}_{#1}}

\newcommand{\DiscMod}[1]{\mathcal{D}_{#1}}
\newcommand{\IrrMod}[1]{\mathcal{L}_{#1}}
\newcommand{\TypMod}[1]{\mathcal{E}_{#1}}
\newcommand{\StagMod}[1]{\mathcal{S}_{#1}}

\newcommand{\conjaut}{\mathsf{w}} % The conjugation automorphism
\newcommand{\sfaut}{\sigma} % The spectral flow automorphism
\newcommand{\conjmod}[1]{\tfunc{\conjaut}{#1}} % Conjugate a module
\newcommand{\sfmod}[2]{\tfunc{\sfaut^{#1}}{#2}} % Apply spectral flow #1 times to #2

\newcommand{\SLA}[2]{\alg{#1} \left( #2 \right)}
\newcommand{\AKMA}[2]{\affine{\alg{#1}} \left( #2 \right)}
\newcommand{\AKMSA}[3]{\affine{\alg{#1}} \left( #2 \middle\vert #3 \right)}
\newcommand{\SLG}[2]{\grp{#1} \left( #2 \right)}

\newcommand{\WAlg}[1]{\mathbb{W}^{\brac{#1}}}
\newcommand{\WTypMod}[2]{\mathbb{E}_{#1}^{\brac{#2}}}

\newcommand{\ch}[1]{\mathrm{ch} \bigl[ #1 \bigr]}
\newcommand{\fch}[2]{\ch{#1} \left( #2 \right)}

\newcommand{\partfunc}{\mathrm{Z}}

\newcommand{\modS}{\mathsf{S}}
\newcommand{\modT}{\mathsf{T}}
\newcommand{\modarg}[3]{\left( \: #1 \: \middle\vert \: #2 \: \middle\vert \: #3 \: \right)}
\newcommand{\chmodS}{\dot{\mathsf{S}}}
\newcommand{\WmodS}{\mathbb{S}}

\newcommand{\fuse}{\mathbin{\times}}
\newcommand{\chfuse}{\mathbin{\dot{\times}}}
\newcommand{\fuscoeff}[2]{\mathsf{N}_{#1}^{\hphantom{#1} #2}}

\newcommand{\normord}[1]{\mbox{${} : #1 : {}$}} % {} necessary to prevent := or =:

\newcommand{\jth}[1]{\vartheta_{#1}}
\newcommand{\Jth}[2]{\jth{#1} \bigl( #2 \bigr)}

\newcommand{\dses}[3]{0 \longrightarrow #1 \longrightarrow #2 \longrightarrow #3 \longrightarrow 0}
\newcommand{\dres}[5]{\cdots \longrightarrow #5 \longrightarrow #4 \longrightarrow #3 \longrightarrow #2 \longrightarrow #1 \longrightarrow 0}

\newcommand{\eqnref}[1]{Equation~\eqref{#1}}
\newcommand{\eqnDref}[2]{Equations~\eqref{#1} and \eqref{#2}}
\newcommand{\secref}[1]{Section~\ref{#1}}
\newcommand{\secDref}[2]{Sections~\ref{#1} and \ref{#2}}
\newcommand{\appref}[1]{Appendix~\ref{#1}}
\newcommand{\figref}[1]{Figure~\ref{#1}}

\newcommand{\cft}{conformal field theory}
\newcommand{\cfts}{conformal field theories}

\newcommand{\lcft}{logarithmic conformal field theory}
\newcommand{\lcfts}{logarithmic conformal field theories}
\newcommand{\WZW}{Wess-Zumino-Witten}

\newcommand{\hws}{highest weight state}

\newcommand{\lws}{lowest weight state}
\newcommand{\lwss}{lowest weight states}

\newcommand{\hwms}{highest weight modules}

\DeclareMathOperator{\Ind}{Ind}
\DeclareMathOperator{\tr}{tr}
\DeclareMathOperator{\id}{id}

\begin{document}

\title[Modular Data and Verlinde Formulae for Fractional Level WZW Models I]{Modular Data and Verlinde Formulae \\ for Fractional Level WZW Models I}

\author[T Creutzig]{Thomas Creutzig}

\address[T Creutzig]{
Fachbereich Mathematik \\
Technische Universit\"{a}t Darmstadt\\
Schlo\ss{}gartenstra\ss{}e 7\\
64289 Darmstadt\\ Germany
}

\email{tcreutzig@mathematik.tu-darmstadt.de}

\author[D Ridout]{David Ridout}

\address[David Ridout]{
Department of Theoretical Physics \\
Research School of Physics and Engineering;
and
Mathematical Sciences Institute;
Australian National University \\
Canberra, ACT 0200 \\
Australia
}

\email{david.ridout@anu.edu.au}

\thanks{\today}

\begin{abstract}
The modular properties of fractional level $\AKMA{sl}{2}$-theories and, in particular, the application of the Verlinde formula, have a long and checkered history in \cft{}.  Recent advances in \lcft{} have led to the realisation that problems with fractional level models stem from trying to build the theory with an insufficiently rich category of representations.  In particular, the appearance of negative fusion coefficients for admissible highest weight representations is now completely understood.  Here, the modular story for certain fractional level theories is completed.  Modular transformations are derived for the complete set of admissible irreducible representations when the level is $k=-\tfrac{1}{2}$ or $k=-\tfrac{4}{3}$.  The S-matrix data and Verlinde formula are then checked against the known fusion rules with complete agreement.  Finally, an infinite set of modular invariant partition functions is constructed in each case.% and conjectures are made for the structure of the non-chiral state space.
\end{abstract}

\maketitle

\onehalfspacing

\section{Introduction} \label{sec:Intro}

This is the fourth part of a series of articles devoted to a detailed understanding, as \cfts{}, of the fractional level \WZW{} models with symmetry algebra $\AKMA{sl}{2}$.  The first three articles \cite{RidSL208,RidSL210,RidFus10} deal exclusively with the case in which the level of $\AKMA{sl}{2}$ is $k=-\tfrac{1}{2}$.  The first described the minimal (chiral) spectrum that is obtained from the admissible \hwms{} of \cite{KacMod88} by demanding closure under conjugation and fusion, clarified the relation between the chiral algebra $\AKMA{sl}{2}_{-1/2}$ and that of the $\beta \gamma$ ghost system, and solved the long-standing issue of why a na\"{\i}ve application of the Verlinde formula to this model results in negative ``fusion coefficients'' \cite{KohFus88,BerFoc90,MatFra90}.  The second article of this series proved a remarkable relationship between $\AKMA{sl}{2}_{-1/2}$ and the triplet algebra $W \brac{1,2}$ of \cite{KauExt91,KauCur95,GabRat96}.  This was then used to motivate and study a natural extension of the spectrum.  The third article tackled the difficult task of computing the fusion rules for this extended spectrum, proving that this generated reducible yet indecomposable modules of the type called ``staggered'' in the Virasoro literature \cite{RohRed96,RidSta09}.  The present article deals again primarily with the level $k=-\tfrac{1}{2}$, though the case $k=-\tfrac{4}{3}$, previously studied in \cite{GabFus01,Ad,AdaLat09}, is also discussed in depth.  We will report on the generalisations of the results presented here to arbitrary (admissible) fractional level $\AKMA{sl}{2}$-models in a sequel \cite{RidMod2}.

The issue regarding negative fusion coefficients in fractional level theories has a long history.  Such theories were initially proposed \cite{KenInf86} as a speculative generalisation of the theories with $k$ a non-negative integer, the idea being that such generalisations would allow a coset construction of the non-unitary minimal models that naturally generalised that of their unitary cousins \cite{GodUni86}.  Shortly thereafter, it was shown \cite{KacMod88,KacMod88b} that certain levels possess a finite set of \emph{admissible} \hwms{} which carry a representation of the modular group $\SLG{SL}{2 ; \ZZ}$.  With the announcement of Verlinde's formula \cite{VerFus88} for fusion coefficients at around the same time, its application to fractional level fusion rules seemed natural.  However, the negative coefficients \cite{KohFus88,BerFoc90,MatFra90} so derived demanded explanation and did not match those that were deduced using more traditional means \cite{AwaFus92,FeiFus94,AndOpe95,DonVer97,PetFus96,FurAdm97}.  Of course, much was made of the observation \cite{BerFoc90,AwaFus92,RamNew93} that the two approaches to fusion could be reconciled if negative signs for coefficients were interpreted as indicating that the multiplicity applied to modules that were \emph{conjugate} to certain admissible \hwms{}.  However, it seems that no compelling explanation for this interpretation was known.  Because of this, the status of fractional level theories as \cfts{} was regarded as questionable at best and ``intrinsically sick'' \cite{DiFCon97} at worst.

A significant advance in the understanding of fractional level theories was reported in \cite{GabFus01}, where the algebraic approach to fusion that has come to be known as the Nahm-Gaberdiel-Kausch algorithm \cite{NahQua94,GabInd96} was applied to $\AKMA{sl}{2}$ at level $k=-\tfrac{4}{3}$.  There, it was pointed out that the sets of admissible \hwms{} and conjugates of these modules that had been considered in the literature were \emph{not} closed under fusion.  Rather, fusion also generates modules whose conformal dimensions are unbounded both above and below, as well as modules on which the Virasoro mode $L_0$ acts non-diagonalisably.  The presence of the former class of modules should actually have been expected, as they arise from applying the familiar symmetries known as spectral flow transformations.  The presence of the latter class indicates that $k=-\tfrac{4}{3}$ fractional level models are \lcfts{}.  The general validity of these conclusions was subsequently tested on $\AKMA{sl}{2}$ at $k=-\tfrac{1}{2}$ in \cite{LesSU202,LesLog04} using free field methods.  There, some evidence was given supporting the natural prediction that this level also leads to modules with conformal dimensions unbounded from below and logarithmic theories (though it was also claimed that a non-logarithmic theory at this level also exists).  These predictions were eventually confirmed as corollaries of the detailed analysis of \cite{RidSL208,RidSL210,RidFus10} --- we refer the reader to \cite[Sec.~6.1]{RidSL210} for a discussion of this.

We note that neither \cite{GabFus01} nor \cite{LesSU202,LesLog04} tried seriously to address the modular properties of the theories at $k=-\tfrac{4}{3}$ and $k=-\tfrac{1}{2}$, respectively.  As remarked above, \cite{RidSL208} gives a complete understanding of why the Verlinde formula gives negative fusion coefficients when applied to the admissible \hwms{}.  However, this says next to nothing about the infinitely many other irreducible modules in the spectrum (which are not highest weight).  In what follows, we will rectify this shortcoming.  Roughly speaking, the negative coefficients arise because the characters of the infinitely many irreducible modules are not all linearly independent as meromorphic functions:  The map from modules to characters is \emph{infinite}-to-one.  Our approach is simply to reject the interpretation of characters as meromorphic functions, and instead regard them as formal power series (or more precisely, as algebraically-defined distributions).  This allows us to work with a one-to-one map between irreducible modules and characters.  A distributional interpretation of characters is nothing new in non-rational \cft{} (see \cite{MalStr01,BarMod11} for example), however we believe that this is the first time it has been applied to fractional level theories.\footnote{We have recently learned that the idea of a distributional approach to fractional level characters was briefly proposed in \cite{LesLog04}, but was discarded on the grounds that the approach did not seem sufficiently fruitful.}  The resulting modular properties are beautiful.

The article is organised as follows.  We first introduce, in \secref{sec:Review}, our notations and conventions for $\AKMA{sl}{2}_{-1/2}$ and its representations, reviewing what will be needed concerning the  spectrum, fusion and character formulae.  \secref{sec:ModTrans} then considers the modular properties of the irreducible characters, considered as distributions.  The idea here parallels that which was used to study the modular transformations of $\AKMSA{gl}{1}{1}$-modules and their extended algebras in \cite{CreBra08,CreRel11,Alfes:2012pa} and leads to a (projective) representation of $\SLG{SL}{2 ; \ZZ}$ for which the S- and T-matrices are both symmetric and unitary.  Here, ``matrix'' should be understood in a generalised sense in which the entries are indexed by an uncountable set.  These S-matrix ``entries'' are then used to calculate fusion ``coefficients'' (which are actually distributions) using the obvious continuum analogue of the Verlinde formula.  The results are then checked to agree perfectly with the projections of the fusion calculations of \cite{RidFus10} onto the Grothendieck ring of characters.

With the modular properties completely understood, \secref{sec:ModInv} turns to the question of constructing modular invariant partition functions.  We note the invariance of the diagonal and charge-conjugation partition functions before showing that there exists an additional infinite family of modular invariants, one of which is the invariant predicted in \cite{RidSL210} on the basis of the relationship between $\AKMA{sl}{2}_{-1/2}$ and $W \brac{1,2}$.  Following \cite{CreWAl11}, these are then interpreted as simple current invariants arising from diagonal invariants for extended algebras.%  We conclude the section by using the extended algebra formalism to formulate and justify natural conjectures for the structure of the non-chiral state space.

\secref{sec:k=-4/3} then applies the technology we have developed for $\AKMA{sl}{2}_{-1/2}$ to the only other case for which reliable fusion data is known:  $\AKMA{sl}{2}_{-4/3}$.  After again reviewing what is known about theories with this symmetry algebra, we derive the modular transformations of the characters and apply Verlinde.  This time, the results confirm the fusion rules reported in \cite{GabFus01}, but are inconsistent with a conjectured fusion rule stated there.  We correct this and verify that there is again an infinite family of simple current extensions leading to modular invariant partition functions.  One of these may be identified as that which relates $\AKMA{sl}{2}_{-4/3}$ to the $c=-7$ triplet algebra $W\brac{1,3}$ \cite{Ad,CRW}.  The article concludes with two appendices:  The first briefly reviews an important identity that we need in the text.  It follows rather easily from the denominator formula of $\AKMSA{sl}{2}{1}$.  The second is devoted to proving that the structure diagrams which we have derived for certain $k=-\tfrac{4}{3}$ indecomposables actually determine them up to isomorphism.

It is very interesting to note that the extended theories which we have constructed at both $k=-\tfrac{1}{2}$ and $k=-\tfrac{4}{3}$ are rational \lcfts{}, meaning that their spectra each contain only finitely many irreducible modules.  We expect that a detailed analysis of their properties will help to illuminate many of the poorly-understood properties of rational logarithmic theories (see, for instance, \cite{GabLoc99,FucNon04,FeiMod06,PeaGro10,GabMod11}).  It is too tempting not to speculate that these extended theories might have a geometric realisation in terms of the \WZW{} models on $\SLG{SL}{2 ; \RR}$ and its covers.\footnote{$\SLG{SL}{2 ; \RR}$ and its covers are all non-compact three-dimensional Lie groups, hence their third integral cohomology groups vanish.  It follows that the standard \WZW{} action on these groups imposes no quantisation constraints upon the level $k$.}  Since the fundamental group of $\SLG{SL}{2 ; \RR}$ is isomorphic to $\ZZ$ and its centre is isomorphic to $\ZZ_2$, it seems plausible that the diagonal modular invariant describes strings propagating on the simply-connected universal cover (that physicists sometimes refer to as $\mathsf{AdS}_3$) and that the ``smallest'' of the rational simple current invariants corresponds to strings on $\SLG{SL}{2 ; \RR}$, or perhaps $\SLG{PSL}{2 ; \RR}$.  It would be interesting to pursue such realisations in earnest.

In general, we believe that fractional level theories are excellent toy models for learning about logarithmic, and even non-rational, theories in general.  Their relatively accessible algebras, coupled with the well-behaved modular properties derived here, only bolster our belief that these theories should be regarded as fundamental building blocks for bulk and boundary \lcfts{}, much as their non-negative integer level cousins are for rational \cfts{}.  It is worth emphasising that having a working Verlinde formula, assuming of course that it can be shown to be valid, enables one to easily deduce an enormous amount of information about fusion rules that is otherwise extremely difficult to obtain.  In particular, we can obtain the character of a fusion product.  This then can be used to decide how more detailed methods, the Nahm-Gaberdiel-Kausch fusion algorithm in particular, should be applied and when they should be terminated.  We hope to report on the results of combining these methods in the future.

\section{Background} \label{sec:Review}

\subsection{Algebraic Preliminaries}

This section serves to review the results of \cite{RidSL208,RidSL210,RidFus10} concerning theories with $\AKMA{sl}{2}_k$-symmetry, specialising to $k=-\tfrac{1}{2}$ when necessary.  We fix a basis $\set{e,h,f}$ of $\SLA{sl}{2}$ so that the non-vanishing commutation relations are
\begin{equation}
\comm{h}{e} = 2 e, \qquad \comm{e}{f} = -h, \qquad \comm{h}{f} = -2 f.
\end{equation}
This basis is preferred because it is compatible with a triangular decomposition respecting the $\SLA{sl}{2 ; \RR}$ adjoint, $e^{\dag} = f$ and $h^{\dag} = h$, the latter being required to realise the $\beta \gamma$ ghost system as a simple current extension of $\AKMA{sl}{2}_{-1/2}$.  The Killing form is given in this basis by
\begin{equation}
\killing{h}{h} = 2, \qquad \killing{e}{f} = -1,
\end{equation}
with all other combinations giving zero.

The non-vanishing commutation relations of $\AKMA{sl}{2}$ are therefore
\begin{equation}
\comm{h_m}{e_n} = 2 e_{m+n}, \quad \comm{h_m}{h_n} = 2m \delta_{m+n,0} k, \quad \comm{e_m}{f_n} = -h_{m+n} - m \delta_{m+n,0} k, \quad \comm{h_m}{f_n} = -2 f_{m+n}.
\end{equation}
As we will habitually identify the level $k$ with its common eigenvalue on the modules being considered, it is convenient to denote this algebra by $\AKMA{sl}{2}_k$ when we wish to emphasise the value that the level is taking.  The energy-momentum tensor is now uniquely specified by requiring that the current fields $\func{e}{z}$, $\func{h}{z}$ and $\func{f}{z}$ are conformal primaries of dimension $1$.  It is given by
\begin{equation} \label{eqnDefT}
\func{T}{z} = \frac{1}{2 \brac{k+2}} \brac{\frac{1}{2} \normord{\func{h}{z} \func{h}{z}} - \normord{\func{e}{z} \func{f}{z}} - \normord{\func{f}{z} \func{e}{z}}}
\end{equation}
and the central charge is $c = 3k / \brac{k+2}$, hence $c=-1$ when $k=-\tfrac{1}{2}$.  The modes of the energy-momentum tensor are denoted, as usual, by $L_n$.

The automorphisms of $\AKMA{sl}{2}$ which preserve the span of the zero-modes $h_0$, $k$ and $L_0$ are generated by the conjugation automorphism $\conjaut$ and the spectral flow automorphism $\sfaut$.  The first may be identified with the non-trivial Weyl reflection of $\SLA{sl}{2}$ and the latter with a square root of the generator of the translation subgroup of the affine Weyl group of $\AKMA{sl}{2}$ (it corresponds to translation by the dual of the simple root of $\SLA{sl}{2}$).  These automorphisms act as follows (both leave $k$ invariant):
\begin{equation}
\begin{aligned}
\func{\conjaut}{e_n} &= f_n, \\ 
\func{\sfaut}{e_n} &= e_{n-1},
\end{aligned}
\qquad
\begin{aligned}
\func{\conjaut}{h_n} &= -h_n, \\
\func{\sfaut}{h_n} &= h_n - \delta_{n,0} k,
\end{aligned}
\qquad
\begin{aligned}
\func{\conjaut}{f_n} &= e_n, \\
\func{\sfaut}{f_n} &= f_{n+1},
\end{aligned}
\qquad
\begin{aligned}
\func{\conjaut}{L_0} &= L_0, \\
\func{\sfaut}{L_0} &= L_0 - \tfrac{1}{2} h_0 + \tfrac{1}{4} k.
\end{aligned}
\end{equation}
Note that $\conjaut \sfaut = \sfaut^{-1} \conjaut$.  Our interest in these automorphisms stems from the fact that they may be used to ``twist'' the action of $\AKMA{sl}{2}$ on a module $\mathcal{M}$, thereby obtaining new modules $\tfunc{\conjaut^*}{\mathcal{M}}$ and $\func{\sfaut^*}{\mathcal{M}}$.  The first is just the module conjugate to $\mathcal{M}$.  We shall refer to the second (and its iterates under repeated twists) as being ``spectrally-flowed'' or just ``twisted''.  Explicitly, the twisted algebra action defining $\tfunc{\conjaut^*}{\mathcal{M}}$ and $\func{\sfaut^*}{\mathcal{M}}$ is given by
\begin{equation}
J \cdot \tfunc{\conjaut^*}{\ket{v}} = \func{\conjaut^*}{\func{\conjaut^{-1}}{J} \ket{v}}, \qquad J \cdot \tfunc{\sfaut^*}{\ket{v}} = \func{\sfaut^*}{\func{\sfaut^{-1}}{J} \ket{v}} \qquad \text{($J \in \AKMA{sl}{2}$).}
\end{equation}
In what follows, we will not bother with the superscript ``$*$'' which distinguishes the algebra automorphisms and the induced maps between modules.  Which is meant will be clear from the context.

\subsection{Representation Theory} \label{sec:RepTheory}

Our chosen triangular decomposition of $\SLA{sl}{2}$ (with $e$ an annihilation operator and $f$ a creation operator) lifts to a decomposition of $\AKMA{sl}{2}$ that allows one to define the usual notion of a Verma module and its irreducible quotient.  However, these turn out to be insufficient for the field-theoretic applications at hand.  We will instead consider the generalised triangular decomposition
\begin{equation}
\AKMA{sl}{2} = \alg{g}^- \oplus \alg{g}^0 \oplus \alg{g}^+,
\end{equation}
where $\alg{g}^{\pm}$ is the subalgebra of $\AKMA{sl}{2}$ generated by the $e_n$, $h_n$ and $f_n$ with $\pm n > 0$ and $\alg{g}^0$ is the subalgebra spanned by $e_0$, $h_0$, $f_0$ and $k$.  Note that $\alg{g}^0$ is isomorphic to $\SLA{gl}{2}$.  An $\SLA{sl}{2}$-module $\finite{\mathcal{M}}$ may then be regarded as a $\alg{g}^0$-module by letting $k$ act as a multiple of the identity (this multiple is also denoted by $k$) and then as a $\alg{g}^0 \oplus \alg{g}^+$-module by letting $\alg{g}^+$ act trivially.  From this, we can apply the induced module construction to obtain an $\AKMA{sl}{2}_k$-module $\mathcal{M} \uparrow {}$:
\begin{equation}
\mathcal{M} \uparrow {} = \Ind_{\alg{g}^0 \oplus \alg{g}^+}^{\AKMA{sl}{2}_k} \finite{\mathcal{M}}.
\end{equation}
If $\finite{\mathcal{M}}$ is an irreducible $\SLA{sl}{2}$-module, then the induced module $\mathcal{M} \uparrow {}$ will have a unique maximal submodule $\mathcal{J}$.  We shall denote the irreducible $\AKMA{sl}{2}$-module $\mathcal{M} \uparrow {} / \mathcal{J}$ by $\mathcal{M}$ and refer to it as the affinisation of $\finite{\mathcal{M}}$.  We further remark that one can also obtain twisted versions $\sfmod{\ell}{\mathcal{M}}$ by choosing a different $\SLA{sl}{2}$-subalgebra to commence the induction procedure.

The classification of irreducible $\SLA{sl}{2}$-modules results in the following list.
\begin{itemize}
\item[$\FinIrrMod{\lambda}$:] There is a \hws{} of weight $\lambda \in \NN$ and a \lws{} of weight $-\lambda$.
\item[$\FinDiscMod{\lambda}^+$:] There is a \hws{} of weight $\lambda \notin \NN$ and no \lws{}.
\item[$\FinDiscMod{\lambda}^-$:] There is a \lws{} of weight $\lambda \notin -\NN$ and no \hws{}.
\item[$\FinTypMod{\lambda,\Delta}$:] There are no highest or \lwss{}.  $\lambda$ denotes the common weight of the states $\bmod 2$ and $\Delta$ the eigenvalue of the Casimir.
\end{itemize}
The multiplicities of the weight spaces ($h_0$-eigenspaces) are always one.  We remark that the weight of a highest or \lws{} completely determines the eigenvalue of the Casimir, but this eigenvalue is free in the absence of a highest or \lws{}.  However, when $\lambda$ and $\Delta$ satisfy the relation appropriate to having a highest or \lws{}, the module denoted above by $\FinTypMod{\lambda,\Delta}$ will no longer be irreducible.  We will not attempt to spell out the precise irreducibility conditions for these modules, though it is not difficult, but will return to this point later.

We will call an $\AKMA{sl}{2}_k$-module \emph{admissible} if it defines a representation of the vertex algebra whose space of states is the irreducible $\AKMA{sl}{2}_k$-module $\IrrMod{0}$.  This generalises the original notion of admissibility \cite{KacMod88} beyond \hwms{} and category $\mathcal{O}$ (see \cite{AdaVer95}).  Indeed, physical consistency requires us to enlarge our module category significantly.  We therefore propose a relaxation of the axioms of category $\mathcal{O}$ so that the objects satisfy:
\begin{enumerate}
\item Each $\AKMA{sl}{2}_k$-module $\mathcal{M}$ is finitely generated.
\item $h_0$ acts semisimply on $\mathcal{M}$ (though $L_0$ need not).
\item Given any $\ket{v} \in \mathcal{M}$, there exists $N>0$ such that $J_n \ket{v} = 0$ for all $J \in \SLA{sl}{2}$ and $n>N$. \label{it:NewAxiom}
\end{enumerate}
The morphisms are the usual module homomorphisms.  We remark that (\ref{it:NewAxiom}) is where our category differs from category $\mathcal{O}$ (there, the space obtained by acting on $\ket{v}$ with arbitrary linear combinations of monomials in the positive modes is required to be finite-dimensional for each $\ket{v}$).  This also generalises the twisted relaxed category considered in \cite{FeiEqu98} which is not closed under fusion (it does not admit the staggered modules that we shall introduce in \secref{sec:Fusion-1/2}).

Specialising now to $k=-\tfrac{1}{2}$, the admissible irreducible $\AKMA{sl}{2}_{-1/2}$-modules from this relaxed category fall into two countably-infinite families and one uncountably-infinite family:\footnote{We mention that we will always choose $\left( -1,1 \right]$ as the fundamental domain of $\RR / 2 \ZZ$.}
\begin{equation*}
\sfmod{\ell}{\IrrMod{0}}, \quad \sfmod{\ell}{\IrrMod{1}}, \quad \sfmod{\ell}{\TypMod{\lambda}} \qquad \text{($\ell \in \ZZ$, $\lambda \in \left( -1 , 1 \right] \setminus \set{\pm \tfrac{1}{2}}$).}
\end{equation*}
Here, $\IrrMod{\lambda}$ denotes the affinisation of the $\SLA{sl}{2}$-module $\FinIrrMod{\lambda}$ and $\TypMod{\lambda}$ denotes the affinisation of $\FinTypMod{\lambda,\Delta}$, where the Casimir eigenvalue $\Delta$ is chosen so that the minimal conformal dimension among the states of $\TypMod{\lambda}$ is $-\tfrac{1}{8}$.  We will discuss the excluded case $\lambda \in \set{\pm \tfrac{1}{2}}$ in some detail in the next section.  Finally, note also the identifications
\begin{equation}
\sfmod{}{\IrrMod{0}} = \DiscMod{-1/2}^+, \qquad \sfmod{-1}{\IrrMod{0}} = \DiscMod{1/2}^-, \qquad \sfmod{}{\IrrMod{1}} = \DiscMod{-3/2}^+, \qquad \sfmod{-1}{\IrrMod{1}} = \DiscMod{3/2}^-
\end{equation}
which relate the twisted modules $\sfmod{\pm 1}{\IrrMod{0}}$ and $\sfmod{\pm 1}{\IrrMod{1}}$ to the affinisations of the $\SLA{sl}{2}$-modules $\FinDiscMod{\pm 1/2}^{\mp}$ and $\FinDiscMod{\pm 3/2}^{\mp}$.  The irreducible spectrum is illustrated schematically in \figref{fig:Spec} for convenience.

{
\psfrag{L0}[][]{$\IrrMod{0}$}
\psfrag{L1}[][]{$\IrrMod{1}$}
\psfrag{E0}[][]{$\TypMod{\lambda}$}
\psfrag{La}[][]{$\DiscMod{-1/2}^+$}
\psfrag{Lb}[][]{$\DiscMod{-3/2}^+$}
\psfrag{La*}[][]{$\DiscMod{1/2}^-$}
\psfrag{Lb*}[][]{$\DiscMod{3/2}^-$}
\psfrag{g}[][]{$\sfaut$}
\psfrag{00}[][]{$\scriptstyle \brac{0,0}$}
\psfrag{aa}[][]{$\scriptstyle \tbrac{-\tfrac{1}{2},-\tfrac{1}{8}}$}
\psfrag{bb}[][]{$\scriptstyle \tbrac{\tfrac{1}{2},-\tfrac{1}{8}}$}
\psfrag{cc}[][]{$\scriptstyle \tbrac{-1,-\tfrac{1}{2}}$}
\psfrag{dd}[][]{$\scriptstyle \tbrac{1,-\tfrac{1}{2}}$}
\psfrag{ee}[][]{$\scriptstyle \tbrac{1,\tfrac{1}{2}}$}
\psfrag{ff}[][]{$\scriptstyle \tbrac{-1,\tfrac{1}{2}}$}
\psfrag{gg}[][]{$\scriptstyle \tbrac{-\tfrac{3}{2},-\tfrac{1}{8}}$}
\psfrag{hh}[][]{$\scriptstyle \tbrac{\tfrac{1}{2},\tfrac{7}{8}}$}
\psfrag{ii}[][]{$\scriptstyle \tbrac{\tfrac{3}{2},-\tfrac{1}{8}}$}
\psfrag{jj}[][]{$\scriptstyle \tbrac{-\tfrac{1}{2},\tfrac{7}{8}}$}
\psfrag{kk}[][]{$\scriptstyle \brac{-2,-1}$}
\psfrag{ll}[][]{$\scriptstyle \brac{0,1}$}
\psfrag{mm}[][]{$\scriptstyle \brac{2,-1}$}
\psfrag{0e}[][]{$\scriptstyle \tbrac{\lambda,-\tfrac{1}{8}}$}
\psfrag{aq}[][]{$\scriptstyle \tbrac{\lambda-\tfrac{1}{2},\tfrac{\lambda}{2}-\tfrac{1}{4}}$}
\psfrag{cq}[][]{$\scriptstyle \tbrac{\lambda-1,\lambda-\tfrac{5}{8}}$}
\psfrag{bq}[][]{$\scriptstyle \tbrac{\lambda+\tfrac{1}{2},-\tfrac{\lambda}{2}-\tfrac{1}{4}}$}
\psfrag{dq}[][]{$\scriptstyle \tbrac{\lambda+1,-\lambda-\tfrac{5}{8}}$}
\begin{figure}
\begin{center}
\includegraphics[width=\textwidth]{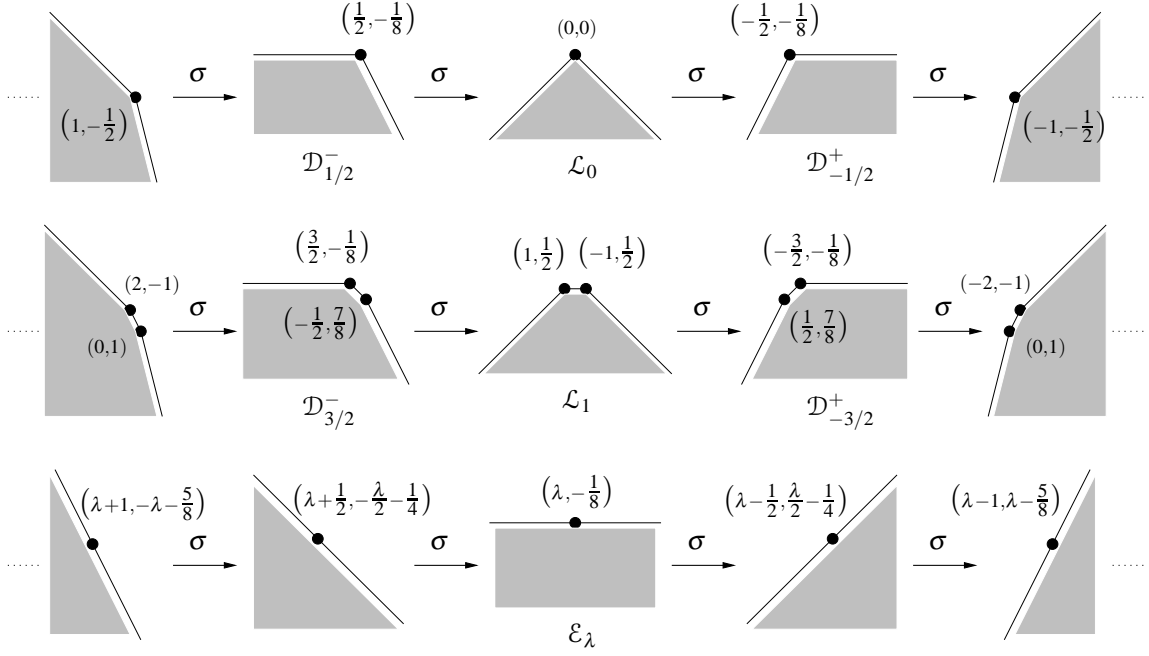}
\caption{Depictions of the admissible irreducible $\AKMA{sl}{2}_{-1/2}$-modules.  Each labelled state declares its $\func{\alg{sl}}{2}$-weight and conformal dimension (in that order).  Conformal dimensions increase from top to bottom and $\SLA{sl}{2}$-weights increase from right to left.} \label{fig:Spec}
\end{center}
\end{figure}
}

\subsection{Fusion} \label{sec:Fusion-1/2}

The fusion rules of the admissible $k=-\tfrac{1}{2}$ irreducibles are known, up to a conjectured relation concerning the interplay between spectral flow and fusion, assumed to be valid for all modules $\mathcal{M}$ and $\mathcal{N}$:
\begin{equation} \label{eqn:FusionAssumption}
\sfmod{\ell_1}{\mathcal{M}} \fuse \sfmod{\ell_2}{\mathcal{N}} = \sfmod{\ell_1 + \ell_2}{\mathcal{M} \fuse \mathcal{N}}.
\end{equation}
We know of no proof for this relation despite much evidence in its favour.  Assuming its validity, we can restrict the fusion rules to the ``untwisted'' sector, where they take the following form:
\begin{equation} \label{eqn:Fusion1}
\begin{gathered}
\IrrMod{\lambda} \fuse \IrrMod{\mu} = \IrrMod{\lambda + \mu}, \qquad 
\IrrMod{\lambda} \fuse \TypMod{\mu} = \TypMod{\lambda + \mu}, \\
\TypMod{\lambda} \fuse \TypMod{\mu} = 
\begin{cases}
\StagMod{\lambda + \mu} & \text{if $\lambda + \mu \in \ZZ$,} \\
\sfmod{}{\TypMod{\lambda + \mu + 1/2}} \oplus \sfmod{-1}{\TypMod{\lambda + \mu - 1/2}} & \text{otherwise.}
\end{cases}
\end{gathered}
\end{equation}
Here, and in what follows, the addition of weights labelling modules is always taken $\bmod 2$.  This shows that fusion does not close on the untwisted sector and that fusion products need not be completely reducible.  Indeed, fusion generates two additional indecomposable modules $\StagMod{0}$ and $\StagMod{1}$ which are uniquely specified by their structure diagrams:
\begin{center}
\parbox[c]{0.7\textwidth}{
\begin{tikzpicture}[thick,
	nom/.style={circle,draw=black!20,fill=black!20,inner sep=1pt}
	]
\node (top0) at (0,1.5) [] {$\IrrMod{0}$};
\node (left0) at (-1.5,0) [] {$\sfmod{-2}{\IrrMod{1}}$};
\node (right0) at (1.5,0) [] {$\sfmod{2}{\IrrMod{1}}$};
\node (bot0) at (0,-1.5) [] {$\IrrMod{0}$};
\node (top1) at (6,1.5) [] {$\IrrMod{1}$};
\node (left1) at (4.5,0) [] {$\sfmod{-2}{\IrrMod{0}}$};
\node (right1) at (7.5,0) [] {$\sfmod{2}{\IrrMod{0}}$};
\node (bot1) at (6,-1.5) [] {$\IrrMod{1}$};
\node at (0,0) [nom] {$\StagMod{0}$};
\node at (6,0) [nom] {$\StagMod{1}$};
\draw [->] (top0) -- (left0);
\draw [->] (top0) -- (right0);
\draw [->] (left0) -- (bot0);
\draw [->] (right0) -- (bot0);
\draw [->] (top1) -- (left1);
\draw [->] (top1) -- (right1);
\draw [->] (left1) -- (bot1);
\draw [->] (right1) -- (bot1);
\end{tikzpicture}
}.
\end{center}
These indecomposables are the $\AKMA{sl}{2}_{-1/2}$-analogues of the staggered modules studied in \cite{RidSta09} for the Virasoro algebra.  Indeed, $L_0$ acts non-diagonalisably on both $\StagMod{0}$ and $\StagMod{1}$, with Jordan cells of rank at most $2$.  These modules are responsible for the logarithmic nature of \cfts{} with $\AKMA{sl}{2}_{-1/2}$ as chiral symmetry algebra.

Diagrams such as these are to be interpreted as annotated versions of the Loewy diagrams widely used in representation theory to indicate how the composition factors of an indecomposable module are ``glued together''.  More precisely, they visualise a socle series of the $\StagMod{\lambda}$:
\begin{equation}
\begin{gathered}
0 = \StagMod{\lambda}^{\brac{0}} \subset \StagMod{\lambda}^{\brac{1}} \subset \StagMod{\lambda}^{\brac{2}} \subset \StagMod{\lambda}^{\brac{3}} = \StagMod{\lambda}, \\
\StagMod{\lambda}^{\brac{1}} / \StagMod{\lambda}^{\brac{0}} \cong \IrrMod{\lambda}, \qquad 
\StagMod{\lambda}^{\brac{2}} / \StagMod{\lambda}^{\brac{1}} \cong \sfmod{-2}{\IrrMod{\lambda+1}} \oplus \sfmod{2}{\IrrMod{\lambda+1}}, \qquad 
\StagMod{\lambda}^{\brac{3}} / \StagMod{\lambda}^{\brac{2}} \cong \IrrMod{\lambda}.
\end{gathered}
\end{equation}
This is a series whose consecutive quotients are completely reducible and are maximal in this respect.  The fusion rules involving the $\StagMod{\lambda}$ take the form
\begin{equation} \label{eqn:Fusion2}
\IrrMod{\lambda} \fuse \StagMod{\mu} = \StagMod{\lambda + \mu}, \qquad 
\begin{aligned}
\TypMod{\lambda} \fuse \StagMod{\mu} &= \sfmod{-2}{\TypMod{\lambda + \mu + 1}} \oplus 2 \: \TypMod{\lambda + \mu} \oplus \sfmod{2}{\TypMod{\lambda + \mu + 1}}, \\
\StagMod{\lambda} \fuse \StagMod{\mu} &= \sfmod{-2}{\StagMod{\lambda + \mu + 1}} \oplus 2 \: \StagMod{\lambda + \mu} \oplus \sfmod{2}{\StagMod{\lambda + \mu + 1}}.
\end{aligned}
\end{equation}
Those of the twisted modules $\sfmod{\ell}{\StagMod{\lambda}}$ then follow from \eqref{eqn:FusionAssumption}.

We will define the fusion ring as the free abelian group generated by the indecomposables $\sfmod{\ell}{\IrrMod{0}}$, $\sfmod{\ell}{\IrrMod{1}}$, $\sfmod{\ell}{\TypMod{\lambda}}$, $\sfmod{\ell}{\StagMod{0}}$ and $\sfmod{\ell}{\StagMod{1}}$, equipped with the fusion product of \eqref{eqn:Fusion1} and \eqref{eqn:Fusion2}, extended by \eqref{eqn:FusionAssumption}.  There are of course many more indecomposable admissibles beyond those discussed above, as may be seen from the various submodules and quotients of the $\sfmod{\ell}{\StagMod{\lambda}}$.  We will only need to consider four additional families of indecomposables, specifically, the following four submodules of $\sfmod{\pm 1}{\StagMod{\lambda}}$ and their images under spectral flow:
\begin{center}
\parbox[c]{0.9\textwidth}{
\begin{tikzpicture}[thick,
	nom/.style={circle,draw=black!20,fill=black!20,inner sep=1pt}
	]
\node (t0) at (1,2) [] {$\sfmod{-1}{\IrrMod{1}} = \DiscMod{3/2}^-$};
\node (l0) at (0,1) [nom] {$\TypMod{-1/2}^+$};
\node (b0) at (1,0) [] {$\sfmod{}{\IrrMod{0}} = \DiscMod{-1/2}^+$};
\draw [->] (t0) -- (b0);
\node (t1) at (4.5,2) [] {$\sfmod{}{\IrrMod{1}} = \DiscMod{-3/2}^+$};
\node (l1) at (3.5,1) [nom] {$\TypMod{+1/2}^-$};
\node (b1) at (4.5,0) [] {$\sfmod{-1}{\IrrMod{0}} = \DiscMod{1/2}^-$};
\draw [->] (t1) -- (b1);
\node (t2) at (8,2) [] {$\sfmod{-1}{\IrrMod{0}} = \DiscMod{1/2}^-$};
\node (l2) at (7,1) [nom] {$\TypMod{+1/2}^+$};
\node (b2) at (8,0) [] {$\sfmod{}{\IrrMod{1}} = \DiscMod{-3/2}^+$};
\draw [->] (t2) -- (b2);
\node (t3) at (11.5,2) [] {$\sfmod{}{\IrrMod{0}} = \DiscMod{-1/2}^+$};
\node (l3) at (10.5,1) [nom] {$\TypMod{-1/2}^-$};
\node (b3) at (11.5,0) [] {$\sfmod{-1}{\IrrMod{1}} = \DiscMod{3/2}^-$};
\draw [->] (t3) -- (b3);
\end{tikzpicture}
}.
\end{center}
Despite their reducibility, it is clear that these four modules share much in common with the irreducibles $\TypMod{\lambda}$ with $\lambda \notin \ZZ + \tfrac{1}{2}$.  In particular, the minimal conformal dimension of the states of the $\TypMod{\pm 1/2}^{\pm}$ is $-\tfrac{1}{8}$, this space of states of minimal conformal dimension has weights which are unbounded in both directions, and the multiplicity of these minimal weight spaces is always $1$.  Indeed, we will see in \secref{sec:ModAtyp} that the characters of the $\TypMod{\pm 1/2}^{\pm}$ are given by substituting $\lambda = \pm \tfrac{1}{2}$ into the character formula for the irreducible $\TypMod{\lambda}$.  For this reason, we will sometimes refer to the $\TypMod{\lambda}$ and $\TypMod{\pm 1/2}^{\pm}$ (as well as their images under spectral flow) as being $\TypMod{}$-\emph{type modules}.  We remark that the superscript ``$\pm$'' labelling the reducible $\TypMod{}$-type modules is meant to indicate that the indecomposable $\SLA{sl}{2}$-module formed by the minimal conformal dimension states has a highest ($+$) or lowest ($-$) weight state.

While we are introducing terminology, we find it convenient to draw attention to the fact that the $\TypMod{}$-type modules that we have defined are \emph{generically irreducible}, generic here meaning that the weights do not belong to $\ZZ + \tfrac{1}{2}$.  Their images under spectral flow are likewise generically irreducible.  This is reminiscent of the generic irreducibility of (type I) Kac modules in Lie superalgebra theory \cite{KacCha77}, so we will often refer to the irreducible $\TypMod{}$-type modules (and their spectral flow images) as being \emph{typical}.  Similarly, the $\sfmod{\ell}{\TypMod{\pm 1/2}^{\pm}}$ and their subquotients $\sfmod{\ell}{\IrrMod{0}}$ and $\sfmod{\ell}{\IrrMod{1}}$ will be referred to as being \emph{atypical}.

\subsection{Characters} \label{sec:RevChar}

The characters $\fch{\mathcal{M}}{y;z;q} = \tr_{\mathcal{M}} y^k z^{h_0} q^{L_0 - c/24}$ of the admissible \hwms{} are very well known:
\begin{equation} \label{ch:HWMs}
\begin{aligned}
\ch{\IrrMod{0}} &= \frac{y^{-1/2}}{2} \sqbrac{\frac{\func{\eta}{q}}{\Jth{4}{z ; q}} + \frac{\func{\eta}{q}}{\Jth{3}{z ; q}}}, \\
\ch{\IrrMod{1}} &= \frac{y^{-1/2}}{2} \sqbrac{\frac{\func{\eta}{q}}{\Jth{4}{z ; q}} - \frac{\func{\eta}{q}}{\Jth{3}{z ; q}}},
\end{aligned}
\qquad
\begin{aligned}
\ch{\DiscMod{-1/2}^+} &= \frac{y^{-1/2}}{2} \sqbrac{\frac{-\ii \func{\eta}{q}}{\Jth{1}{z ; q}} + \frac{\func{\eta}{q}}{\Jth{2}{z ; q}}}, \\
\ch{\DiscMod{-3/2}^+} &= \frac{y^{-1/2}}{2} \sqbrac{\frac{-\ii \func{\eta}{q}}{\Jth{1}{z ; q}} - \frac{\func{\eta}{q}}{\Jth{2}{z ; q}}}.
\end{aligned}
\end{equation}
The characters of the other ``twisted'' \hwms{} $\sfmod{\ell}{\IrrMod{\lambda}}$ are then obtained from
\begin{equation} \label{eqn:CharSF}
\fch{\sfmod{\ell}{\mathcal{M}}}{y;z;q} = \fch{\mathcal{M}}{y z^{\ell} q^{\ell^2 / 4} ; z q^{\ell / 2} ; q}.
\end{equation}
In doing so, one notices that the characters obtained in this manner are not all linearly independent.  More precisely, the periodicity of the Jacobi theta functions leads to the relations
\begin{equation} \label{eqn:CharDependencies}
\ch{\sfmod{\ell - 1}{\IrrMod{0}}} + \ch{\sfmod{\ell + 1}{\IrrMod{1}}} = 
\ch{\sfmod{\ell - 1}{\IrrMod{1}}} + \ch{\sfmod{\ell + 1}{\IrrMod{0}}} = 0.
\end{equation}
There are therefore only four linearly independent characters among the $\ch{\sfmod{\ell}{\IrrMod{\lambda}}}$, which we may take to be those of the admissible \hwms{} \eqref{ch:HWMs}.

This linear dependence of characters is not matched by corresponding isomorphisms between modules.  Rather \cite{LesSU202}, the characters must be distinguished, as power series in $y$, $z$ and $q$, through their natural domain of convergence.  More precisely, the formal power series that keep track of the (graded) dimensions of the weight space will only converge to the characters \eqref{ch:HWMs} for certain $q$ and $z$.  For $\sfmod{\ell}{\IrrMod{\lambda}}$, this region of convergence is
\begin{equation} \label{AnnulusOfConvergence}
\abs{q} < 1, \qquad \abs{q}^{-\brac{\ell - 1}} < \abs{z}^2 < \abs{q}^{-\brac{\ell + 1}}.
\end{equation}
We emphasise that the linear dependencies \eqref{eqn:CharDependencies} involve characters whose natural convergence domains are \emph{disjoint}, hence these must be understood as relations among the meromorphic extensions \eqref{ch:HWMs} of the characters to the $z$-plane.  We must therefore be careful to distinguish between characters as formal power series and characters as meromorphic functions in what follows.  This distinction makes it clear why the linear dependencies of the meromorphic extensions of the characters do not reflect isomorphisms of (irreducible) modules.  However, modular transformations do not preserve the natural annuli of convergence, so one might think that one is forced to use such extensions to consistently investigate modular properties.

The characters of the irreducible $\TypMod{\lambda}$ (the typical modules) and their twisted cousins are less well known and even more problematic:
\begin{equation} \label{ch:Typs}
\ch{\TypMod{\lambda}} = \frac{y^{-1/2} z^{\lambda}}{\func{\eta}{q}^2} \sum_{n \in \ZZ} z^{2n}.
\end{equation}
This clearly converges nowhere in the $z$-plane.  However, this character formula also applies to the atypical indecomposable modules $\TypMod{\pm 1/2}^{\pm}$, a fact that we shall prove in \secref{sec:ModAtyp}.  These atypical characters are alternatively given, by definition, as the sum of the characters of $\sfmod{\pm 1}{\IrrMod{0}}$ and $\sfmod{\mp 1}{\IrrMod{1}}$.  By \eqref{eqn:CharDependencies}, this sum \emph{vanishes} upon meromorphically extending these characters to the full $z$-plane.  One therefore concludes that the ``meromorphic extensions'' of the characters of the $\TypMod{\pm 1/2}^{\pm}$ are identically $0$.  The same argument applies to the atypical indecomposables $\StagMod{0}$ and $\StagMod{1}$, with the same conclusion.  From here, it seems plausible that the same should be true for the meromorphic extensions of the characters of the typical $\TypMod{\lambda}$ as well.

This proposal has a nicer interpretation at the level of the fusion ring which makes its consistency manifest.  Instead of merely declaring that various (linear combinations of) characters should be set to zero, we assert that we wish to study the quotient of the fusion ring by the \emph{ideal} generated by the modules
\begin{equation}
\sfmod{\ell - 1}{\IrrMod{0}} \oplus \sfmod{\ell + 1}{\IrrMod{1}}, \qquad
\sfmod{\ell - 1}{\IrrMod{1}} \oplus \sfmod{\ell + 1}{\IrrMod{0}}, \qquad
\sfmod{\ell}{\TypMod{\lambda}}, \qquad 
\sfmod{\ell}{\StagMod{0}}, \qquad \sfmod{\ell}{\StagMod{1}}
\end{equation}
whose characters we would like to set to zero.  That this does constitute an ideal is an easy consequence of \eqref{eqn:FusionAssumption}, \eqref{eqn:Fusion1} and \eqref{eqn:Fusion2}.  The quotient ring is free of rank $4$ and we may take the generators to be the equivalence classes of the admissible \hwms{}: $\bigl[ \IrrMod{0} \bigr]$, $\bigl[ \IrrMod{1} \bigr]$, $\bigl[ \DiscMod{-1/2}^+ \bigr]$ and $\bigl[ \DiscMod{-3/2}^+ \bigr]$.  The fusion product descends to the quotient, where we denote it by $\chfuse$, giving the following product:

\begin{center}
\setlength{\extrarowheight}{4pt}
\begin{tabular}{C|CCCC}
\chfuse & \bigl[ \IrrMod{0} \bigr] & \bigl[ \IrrMod{1} \bigr] & \bigl[ \DiscMod{-1/2}^+ \bigr] & \bigl[ \DiscMod{-3/2}^+ \bigr] \\
\hline
\bigl[ \IrrMod{0} \bigr] & \bigl[ \IrrMod{0} \bigr] & \bigl[ \IrrMod{1} \bigr] & \bigl[ \DiscMod{-1/2}^+ \bigr] & \bigl[ \DiscMod{-3/2}^+ \bigr] \\
\bigl[ \IrrMod{1} \bigr] & \bigl[ \IrrMod{1} \bigr] & \bigl[ \IrrMod{0} \bigr] & \bigl[ \DiscMod{-3/2}^+ \bigr] & \bigl[ \DiscMod{-1/2}^+ \bigr] \\
\bigl[ \DiscMod{-1/2}^+ \bigr] & \bigl[ \DiscMod{-1/2}^+ \bigr] & \bigl[ \DiscMod{-3/2}^+ \bigr] & -\bigl[ \IrrMod{1} \bigr] & -\bigl[ \IrrMod{0} \bigr] \\
\bigl[ \DiscMod{-3/2}^+ \bigr] & \bigl[ \DiscMod{-3/2}^+ \bigr] & \bigl[ \DiscMod{-1/2}^+ \bigr] & -\bigl[ \IrrMod{0} \bigr] & -\bigl[ \IrrMod{1} \bigr]
\end{tabular}
\end{center}

\noindent We now have a bijective correspondence between (equivalence classes of) modules and linearly independent meromorphic extensions of characters.  It is not hard to check that the modular S-matrix
\begin{equation}
\chmodS = \frac{1}{2} 
\begin{pmatrix}
1 & -1 & 1 & -1 \\
-1 & 1 & 1 & -1 \\
1 & 1 & \ii & \ii \\
-1 & -1 & \ii & \ii
\end{pmatrix}
\end{equation}
obtained from the latter recovers, through the Verlinde formula, the structure constants of the fusion operation $\chfuse$ on the former.  This, of course, explains how the Verlinde formula is able to give negative integers when applied to the admissible \hwms{} of a fractional level \WZW{} model.

\section{Modular Transformations} \label{sec:ModTrans}

The issue of extending the characters of the admissible modules to meromorphic functions of $\zeta$ is a rather subtle one.  However, we have seen that doing so allows one to reconstruct the structure constants for a certain quotient of the fusion ring from the modular group's action on the (extended) characters.  The price one pays is that the fusion ring quotient is rather small, being generated (as an abelian group) by the four admissible \hwms{}.  Our aim in what follows is to extend the modular group action to the complete set of characters, that is, to the full fusion ring.  More precisely, we will demonstrate that the Verlinde formula reproduces the structure constants of the Grothendieck ring of fusion, defined as the ring obtained from the fusion ring by identifying each indecomposable with the direct sum of its composition factors.\footnote{That the Grothendieck ring is well-defined, meaning that the fusion product descends to the quotient, is not at all obvious.  What one needs to check is that the fusion product is \emph{exact}.  While this can indeed be verified at $k=-\tfrac{1}{2}$ from the known fusion rules, we mention that the irreducibility of the fusion unit ($\IrrMod{0}$) is generally taken as a strong indicator of the exactness of fusion, see \cite{KazTenIV94,TsuTen12}.}  This is the best we can hope for, as the input to a Verlinde formula is the transformation properties of the characters, and characters do not distinguish indecomposable modules from the direct sum of their composition factors.

The key idea that makes such an extension possible is the realisation that one can do better than taking meromorphic extensions of characters for modular considerations:  The correct setting is to extend the characters as \emph{distributions}.  This is not a new idea (see \cite{MalStr01,BarMod11} for example), but applying it to fractional level models gives us the opportunity to analyse the behaviour of such extensions in detail.  In particular, we find that the Verlinde formula relevant to our model is the obvious generalisation of the rational one to a continuous spectrum.  We therefore find this result beautiful, illuminating, and rather satisfactory in the light of the more complicated ``generalised Verlinde formulae'' that have been proposed for other \lcfts{} \cite{FucNon04,FloVer07,GabFro08,SemNot07,PeaGro10}:  There are no troublesome ``$\log q$'' factors and $\tau$-dependent S-matrices to deal with,\footnote{Actually, we will see that the S-matrices we derive do have a weak $\tau$-dependence, but it is entirely contained within the phase $\abs{\tau} / \tau$.  This causes no trouble as such phases always cancel when applying the Verlinde formula or studying non-chiral phenomena such as modular invariants.} and no need to postulate, nor try to interpret, ``pseudo-characters''.  Moreover, we expect that our formalism will lead to a better understanding of these generalised Verlinde formulae.

\subsection{Modular Properties of Typical Characters} \label{sec:ModTyp}

We begin by interpreting \eqnref{ch:Typs} distributionally.  Write $y = \ee^{2 \pi \ii t}$, $z = \ee^{2 \pi \ii \zeta}$ and $q = \ee^{2 \pi \ii \tau}$, so that\footnote{The reader may object at this point to our application of the standard Fourier-theoretic identity $\sum_{n \in \ZZ} \ee^{2 \pi \ii n \zeta} = \sum_{m \in \ZZ} \func{\delta}{\zeta - m}$.  This is indeed problematic because $\zeta$ is complex in general which makes the claim that test functions exist with the appropriate analytic properties rather dubious.  This can be resolved by moving to a purely algebraic characterisation of distribution:  The test functions are the $1$-periodic trigonometric polynomials in $\zeta$ and the pairing with trigonometric power series (the distributions) is effected not by integrating over a period, but by taking the residue, meaning the coefficient of the constant term.}
\begin{equation}
\ch{\TypMod{\lambda}} = \frac{\ee^{-\ii \pi t} \ee^{2 \pi \ii \lambda \zeta}}{\func{\eta}{\tau}^2} \sum_{n \in \ZZ} \ee^{4 \pi \ii n \zeta} = \frac{\ee^{-\ii \pi t} \ee^{2 \pi \ii \lambda \zeta}}{\func{\eta}{\tau}^2} \sum_{m \in \ZZ} \func{\delta}{2 \zeta - m} = \frac{\ee^{-\ii \pi t}}{\func{\eta}{\tau}^2} \sum_{m \in \ZZ} \ee^{\ii \pi m \lambda} \func{\delta}{2 \zeta - m}.
\end{equation}
Applying spectral flow as in \eqref{eqn:CharSF} gives the twisted characters as
\begin{equation} \label{ch:Typ}
\ch{\sfmod{\ell}{\TypMod{\lambda}}} = \frac{\ee^{-\ii \pi t} \ee^{\ii \pi \ell^2 \tau / 4}}{\func{\eta}{\tau}^2} \sum_{m \in \ZZ} \ee^{\ii \pi m \brac{\lambda - \ell / 2}} \func{\delta}{2 \zeta + \ell \tau - m}.
\end{equation}
With the standard action of the modular S- and T-transformations,
\begin{equation}
\modS \colon \modarg{t}{\zeta}{\tau} \longmapsto \modarg{t - \zeta^2 / \tau}{\zeta / \tau}{-1 / \tau}, \qquad
\modT \colon \modarg{t}{\zeta}{\tau} \longmapsto \modarg{t}{\zeta}{\tau + 1}
\end{equation}
(for which $\modS^4 = \brac{\modS \modT}^6 = \id$), we compute
\begin{subequations}
\begin{align}
\modS \set{\ch{\sfmod{\ell}{\TypMod{\lambda}}}} &= \frac{\ee^{-\ii \pi t} \ee^{\ii \pi \zeta^2 / \tau} \ee^{-\ii \pi \ell^2 / 4 \tau}}{\func{\eta}{-1 / \tau}^2} \sum_{m \in \ZZ} \ee^{\ii \pi m \brac{\lambda - \ell / 2}} \func{\delta}{\frac{2 \zeta - \ell - m \tau}{\tau}} \notag \\
&= \frac{\abs{\tau}}{-\ii \tau} \frac{\ee^{-\ii \pi t}}{\func{\eta}{\tau}^2} \sum_{m \in \ZZ} \ee^{\ii \pi m^2 \tau / 4} \ee^{-\ii \pi m \lambda} \func{\delta}{2 \zeta + m \tau - \ell}, \label{eqn:TypS'} \\
\modT \set{\ch{\sfmod{\ell}{\TypMod{\lambda}}}} &= \ee^{\ii \pi \brac{\ell \brac{\lambda - \ell / 4} - 1/6 }} \ch{\sfmod{\ell}{\TypMod{\lambda}}}. \label{eqn:TypT}
\end{align}
\end{subequations}
\eqnref{eqn:TypS'} requires further manipulation.  We claim that the S-transformations of the typical characters take the form
\begin{subequations} \label{eqn:TypS}
\begin{equation} \label{eqn:TypS''}
\modS \set{\ch{\sfmod{\ell}{\TypMod{\lambda}}}} = \sum_{\ell' \in \ZZ} \int_{-1}^1 \modS_{\brac{\ell , \lambda} \brac{\ell' , \lambda'}} \ch{\sfmod{\ell'}{\TypMod{\lambda'}}} \dd \lambda',
\end{equation}
where the ``S-matrix'' elements are given by
\begin{equation} \label{eqn:TypSMat}
\modS_{\brac{\ell , \lambda} \brac{\ell' , \lambda'}} = \frac{1}{2} \frac{\abs{\tau}}{-\ii \tau} \ee^{\ii \pi \brac{\ell \ell' / 2 - \ell \lambda' - \ell' \lambda}}.
\end{equation}
\end{subequations}
Verifying this is straight-forward:  One substitutes \eqref{eqn:TypSMat} and \eqref{ch:Typ} into the right-hand side of \eqref{eqn:TypS''}, performs the $\lambda'$-integration (obtaining $2 \delta_{\ell m}$) then the $m$-summation, and finally relabels $\ell'$ as $m$ to obtain the right-hand side of \eqref{eqn:TypS'}.  This proves that the characters of the typical irreducibles close under modular transformations.

Of course, the S-matrix element \eqref{eqn:TypSMat} has an explicit $\tau$-dependence.  However, we view this as not being a serious problem because this dependence is contained in the phase $\abs{\tau} / \tau = \ee^{-\ii \arg \tau}$.  This phase will cancel in the Verlinde formula as well as when one considers bulk modular invariants.  More abstractly, we see that the typical characters only furnish a \emph{projective} representation\footnote{A projective representation $\pi$ of a group $\grp{G}$ is a representation on a projective vector space $PV$.  When viewed from the perspective of the regular vector space $V$, this means that the representing matrices $\func{\pi}{g}$ need only satisfy the group laws up to a non-zero scalar.  For example, $\func{\pi}{gh} = \omega_{g,h} \func{\pi}{g} \func{\pi}{h}$ for some $\omega_{g,h} \in \CC \setminus \set{0}$.} of the modular group (see \cite{CreRel11} for a related example where the modular representation is only projective).  This may be checked explicitly.  For example, $\modS^2$ has matrix elements
\begin{align}
\sum_{\ell' \in \ZZ} \int_{-1}^1 \modS_{\brac{\ell , \lambda} \brac{\ell' , \lambda'}} \modS_{\brac{\ell' , \lambda'} \brac{\ell'' , \lambda''}} \dd \lambda' 
&= \frac{1}{4} \frac{\abs{\tau}^2}{-\tau^2} \sum_{\ell' \in \ZZ} \int_{-1}^1 \ee^{\ii \pi \brac{\brac{\ell + \ell''} \ell' / 2 - \ell' \brac{\lambda + \lambda''} - \brac{\ell + \ell''} \lambda'}} \dd \lambda' \notag \\
&= -\frac{1}{2} \ee^{-2 \ii \arg \tau} \sum_{\ell' \in \ZZ} \ee^{-\ii \pi \ell' \brac{\lambda + \lambda''}} \delta_{\ell + \ell'' = 0} \notag \\
&= -\ee^{-2 \ii \arg \tau} \delta_{\ell + \ell'' = 0} \func{\delta}{\lambda + \lambda'' = 0 \bmod{2}},
\end{align}
so it follows that $\modS^4 = \ee^{-4 \ii \arg \tau} \id$.  The non-zero phase $\ee^{-4 \ii \arg \tau}$ is a manifestation of the projective nature of the representation.  Here, we use a convenient shorthand for infinite sums of delta functions:  
\begin{equation}
\func{\delta}{\lambda = 0 \bmod{j}} = \sum_{m \in \ZZ} \func{\delta}{\lambda - jm}.
\end{equation}
A similar calculation gives $\brac{\modS \modT}^6 = \ee^{-6 \ii \arg \tau} \id$.  We remark that, up to the phase $-\ee^{-2 \ii \arg \tau}$, $\modS^2$ implements conjugation at the level of chiral characters.  This phase will cancel when we combine chiral characters with their antichiral partners, so $\modS^2$ is precisely conjugation for bulk characters.

Note also that the S-matrix we have computed satisfies
\begin{subequations} \label{eqn:SProps}
\begin{equation}
\modS_{\brac{\ell , \lambda} \brac{\ell' , \lambda'}} = \modS_{\brac{\ell' , \lambda'} \brac{\ell , \lambda}}, \qquad 
\modS_{\brac{\ell , \lambda} \brac{\ell' , \lambda'}} = \modS_{\brac{-\ell , -\lambda} \brac{-\ell' , -\lambda'}}.
\end{equation}
Moreover, a straight-forward calculation shows that it is also unitary (with ${}^{\dag}$ denoting conjugate transpose):
\begin{equation}
\sqbrac{\modS \modS^{\dag}}_{\brac{\ell , \lambda} \brac{\ell'' , \lambda''}} = \delta_{\ell = \ell''} \func{\delta}{\lambda = \lambda'' \bmod{2}}.
\end{equation}
\end{subequations}
These properties give us confidence that the Verlinde formula will work exactly as claimed.  However, before checking this, we must analyse the modular transformations of the atypical characters, in particular, those of the vacuum character $\ch{\IrrMod{0}}$.

\subsection{Modular Properties of Atypical Characters} \label{sec:ModAtyp}

Before studying the modular properties of the characters of the atypical irreducibles $\sfmod{\ell}{\IrrMod{0}}$ and $\sfmod{\ell}{\IrrMod{1}}$, we pause to prove a character identity which we claimed in \secref{sec:RevChar}.  Specifically, we will show that the characters of the atypical indecomposables $\TypMod{\pm 1/2}^{\pm}$ are given by the same formula \eqref{ch:Typs} as those of the typical irreducibles $\TypMod{\lambda}$, hence that the results of \secref{sec:ModTyp} apply to them and their images under spectral flow.  It should be clear from their structures that the characters of the $\TypMod{\pm 1/2}^{\pm}$ do not depend upon the superscript ``$\pm$'', hence this label will sometimes be omitted (in characters) when convenient.  We emphasise that as the atypical characters are non-zero, \eqnref{eqn:CharDependencies} is \emph{incorrect} when characters are treated as distributions rather than as meromorphic functions.

We begin by recalling that the characters of the indecomposables $\TypMod{\pm 1/2}^{\pm}$ are given by the sums of the characters of their composition factors.  The latter are given, as meromorphic functions, by \eqnDref{ch:HWMs}{eqn:CharDependencies}, but we know that the sums will vanish in this setting.  Our strategy will therefore be to expand the composition factor characters in their (disjoint) annuli of convergence \eqref{AnnulusOfConvergence} and then sum them as formal power series in $z$.  This expansion will be carried out using a convenient identity of Kac and Wakimoto which we review in \appref{app:Magic}.

Consider first $\TypMod{1/2}^{\pm}$.  Its character will result from summing the following as formal power series:
\begin{equation} \label{eqn:ToBeSummed}
\begin{aligned}
\ch{\sfmod{-1}{\IrrMod{0}}} &= \frac{y^{-1/2}}{2} \sqbrac{\frac{\ii \func{\eta}{q}}{\Jth{1}{z ; q}} + \frac{\func{\eta}{q}}{\Jth{2}{z ; q}}} \\
\ch{\sfmod{}{\IrrMod{1}}} &= \frac{y^{-1/2}}{2} \sqbrac{\frac{-\ii \func{\eta}{q}}{\Jth{1}{z ; q}} - \frac{\func{\eta}{q}}{\Jth{2}{z ; q}}}
\end{aligned}
\qquad
\begin{aligned}
&\text{($\abs{q} < \abs{z} < 1$),} \vphantom{\sqbrac{\frac{-\ii \func{\eta}{q}}{\Jth{1}{z ; q}}}} \\
&\text{($1 < \abs{z} < \abs{q}^{-1}$).} \vphantom{\sqbrac{\frac{-\ii \func{\eta}{q}}{\Jth{1}{z ; q}}}}
\end{aligned}
\end{equation}
We apply \eqnDref{eqn:Magic1}{eqn:Magic2} to the quotients appearing in these functions in order to deduce the appropriate expansions in $z$.  In this way, we obtain
\begin{equation}
\begin{aligned}
\frac{\ii \func{\eta}{q}}{\Jth{1}{z ; q}} &= \frac{\Jth{1}{u ; q}}{\Jth{1}{uz ; q} \func{\eta}{q}^2} \sum_{n \in \ZZ} \frac{z^n}{1-uq^n} \\
\frac{-\ii \func{\eta}{q}}{\Jth{1}{z ; q}} &= \frac{-\Jth{1}{u ; q}}{\Jth{1}{uz ; q} \func{\eta}{q}^2} \sum_{n \in \ZZ} \frac{z^n uq^n}{1-uq^n}
\end{aligned}
\qquad
\begin{aligned}
&\text{($\abs{q} < \abs{z} < 1$),} \vphantom{\sqbrac{\frac{-\ii \func{\eta}{q}}{\Jth{1}{z ; q}}}} \\
&\text{($1 < \abs{z} < \abs{q}^{-1}$).} \vphantom{\sqbrac{\frac{-\ii \func{\eta}{q}}{\Jth{1}{z ; q}}}}
\end{aligned}
\end{equation}
Treating the right-hand sides as formal power series in $z$, and thus forgetting about the regions of convergence, their sum becomes
\begin{align}
\frac{\Jth{1}{u ; q}}{\Jth{1}{uz ; q} \func{\eta}{q}^2} \sum_{n \in \ZZ} z^n &= \frac{1}{\func{\eta}{q}^2} \sum_{m \in \ZZ} \frac{\Jth{1}{u ; q}}{\Jth{1}{\ee^{2 \pi \ii m} u ; q}} \func{\delta}{\zeta - m} = \frac{1}{\func{\eta}{q}^2} \sum_{m \in \ZZ} \brac{-1}^m \func{\delta}{\zeta - m} \notag \\
&= \frac{z^{1/2}}{\func{\eta}{q}^2} \sum_{n \in \ZZ} z^n,
\end{align}
since $\Jth{1}{\ee^{2 \pi \ii} u ; q} = -\Jth{1}{u ; q}$.  This is the result of summing the $\ii \eta / \jth{1}$ terms in \eqref{eqn:ToBeSummed}.  We can find the sum of the $\eta / \jth{2}$ terms using $\Jth{2}{z ; q} = \Jth{1}{\ee^{\ii \pi} z ; q}$ and thereby compute that
\begin{equation}
\ch{\TypMod{1/2}} = \frac{y^{-1/2} z^{1/2}}{\func{\eta}{q}^2} \sum_{n \in \ZZ} \frac{1}{2} \bigl( 1 + \brac{-1}^n \bigr) z^n = \frac{y^{-1/2} z^{1/2}}{\func{\eta}{q}^2} \sum_{n \in \ZZ} z^{2n},
\end{equation}
as \eqref{ch:Typs} requires.  The result for $\ch{\TypMod{-1/2}}$ follows from an identical analysis (or through conjugation) and applying spectral flow immediately extends this to every $\ch{\sfmod{\ell}{\TypMod{\pm 1/2}}}$.

We can now turn to the modular transformations of the atypical characters.  To attack this problem, it is convenient to summarise the structure of the indecomposables $\sfmod{}{\TypMod{\pm 1/2}^+}$ through the following short exact sequences:
\begin{equation}
\dses{\sfmod{2}{\IrrMod{1}}}{\sfmod{}{\TypMod{1/2}^+}}{\IrrMod{0}}, \qquad 
\dses{\sfmod{2}{\IrrMod{0}}}{\sfmod{}{\TypMod{-1/2}^+}}{\IrrMod{1}}.
\end{equation}
Splicing these sequences iteratively with their appropriately spectrally-flowed versions, we obtain \emph{resolutions} (infinite exact sequences) for the atypical irreducibles $\IrrMod{0}$ and $\IrrMod{1}$:
\begin{equation}
\begin{gathered}
\dres{\IrrMod{0}}{\sfmod{}{\TypMod{1/2}^+}}{\sfmod{3}{\TypMod{-1/2}^+}}{\sfmod{5}{\TypMod{1/2}^+}}{\sfmod{7}{\TypMod{-1/2}^+}}, \\
\dres{\IrrMod{1}}{\sfmod{}{\TypMod{-1/2}^+}}{\sfmod{3}{\TypMod{1/2}^+}}{\sfmod{5}{\TypMod{-1/2}^+}}{\sfmod{7}{\TypMod{1/2}^+}}.
\end{gathered}
\end{equation}
Spectral flow may be applied to obtain resolutions of the other atypical irreducibles.  Exactness now implies the following character identities:
\begin{equation} \label{ch:AtypRes}
\ch{\sfmod{\ell}{\IrrMod{\lambda}}} = \sum_{\ell' = 0}^{\infty} \brac{-1}^{\ell'} \ch{\sfmod{\ell + 2 \ell' + 1}{\TypMod{\lambda + \ell' + 1/2}^+}} \qquad \text{($\lambda = 0,1$).}
\end{equation}
In this way, one expresses the atypical characters in terms of characters of indecomposables which behave, as far as modular transformations are concerned, as if they were typical.  This trick has been used, at the level of characters, in many superalgebra studies, for example \cite{RozSTM93,SalGL106,SalSU207,CreRel11}.  We mention that we could have used the exact sequences describing the $\sfmod{-1}{\TypMod{\pm 1/2}^-}$ to derive resolutions for the $\IrrMod{\lambda}$ (involving negative powers of $\sfaut$).  In either case, the resulting character identities are convergent (as formal power series) as one can easily check that only finitely many $\TypMod{}$-type characters contribute to the multiplicity of any given weight space.

As the characters of the $\sfmod{\ell}{\TypMod{\pm 1/2}^+}$ are given by \eqnref{ch:Typ}, the modular transformations \eqref{eqn:TypT} and \eqref{eqn:TypS} apply to them.  We may therefore check \eqref{ch:AtypRes} by combining it with \eqref{eqn:TypT} to determine the T-transformation of the atypical characters.  This gives
\begin{align} \label{eqn:AtypT}
\modT \set{\ch{\sfmod{\ell}{\IrrMod{\lambda}}}} &= \sum_{\ell' = 0}^{\infty} \brac{-1}^{\ell'} \ee^{\ii \pi \brac{\brac{\ell + 2 \ell' + 1} \brac{\lambda - \ell / 4 + \ell' / 2 + 1/4} - 1/6}} \ch{\sfmod{\ell + 2 \ell' + 1}{\TypMod{\lambda + \ell' + 1/2}^+}} \notag \\
&= \ee^{\ii \pi \brac{\brac{\ell + 1} \lambda - \ell^2 / 4 + 1/12}} \ch{\sfmod{\ell}{\IrrMod{\lambda}}},
\end{align}
which may be checked to agree perfectly with the expected phase $\ee^{2 \pi \ii \brac{\Delta_{\lambda} - c/24}}$ (where $\Delta_{\lambda}$ is the conformal dimension of any state of $\sfmod{\ell}{\IrrMod{\lambda}}$ and $c=-1$ is the central charge).  Combining \eqref{eqn:TypT} and \eqref{ch:AtypRes} then gives the S-transformation of the atypical characters:
\begin{subequations} \label{eqn:AtypS}
\begin{align}
\modS \set{\ch{\sfmod{\ell}{\IrrMod{\lambda}}}} &= \sum_{\ell' = 0}^{\infty} \brac{-1}^{\ell'} \sum_{\ell'' \in \ZZ} \int_{-1}^1 \modS_{\brac{\ell + 2 \ell' + 1 , \lambda + \ell' + 1/2} \brac{\ell'',\lambda''}} \ch{\sfmod{\ell''}{\TypMod{\lambda''}}} \dd \lambda'' \notag \\
&= \frac{1}{2} \frac{\abs{\tau}}{-\ii \tau} \sum_{\ell'' \in \ZZ} \int_{-1}^1 \ee^{\ii \pi \brac{\ell \ell'' / 2 - \ell \lambda'' - \ell'' \lambda - \lambda''}} \ch{\sfmod{\ell''}{\TypMod{\lambda''}}} \sum_{\ell' = 0}^{\infty} \brac{-1}^{\ell'} \ee^{-2 \pi \ii \lambda'' \ell'} \dd \lambda'' \label{eqn:GeomSeries} \\
&= \frac{1}{4} \frac{\abs{\tau}}{-\ii \tau} \sum_{\ell'' \in \ZZ} \int_{-1}^1 \frac{\ee^{\ii \pi \brac{\ell \ell'' / 2 - \ell \lambda'' - \ell'' \lambda}}}{\func{\cos}{\pi \lambda''}} \ch{\sfmod{\ell''}{\TypMod{\lambda''}}} \dd \lambda''.
\end{align}
\end{subequations}
We therefore set (compare with \eqref{eqn:TypSMat})
\begin{equation} \label{eqn:AtypSMat}
\modS_{\overline{\brac{\ell,\lambda}} \brac{\ell',\lambda'}} = \frac{1}{4} \frac{\abs{\tau}}{-\ii \tau} \frac{\ee^{\ii \pi \brac{\ell \ell' / 2 - \ell \lambda' - \ell' \lambda}}}{\func{\cos}{\pi \lambda'}},
\end{equation}
with the overline indicating an irreducible atypical representation.  In particular, we obtain
\begin{equation} \label{eqn:VacSMat}
\modS_{\overline{\brac{0,0}} \brac{\ell,\lambda}} = \frac{1}{4} \frac{\abs{\tau}}{-\ii \tau} \frac{1}{\func{\cos}{\pi \lambda}}
\end{equation}
for the vacuum representation $\IrrMod{0}$.

We remark that we will make no attempt to define S-matrix ``entries'' between two atypical representations as well ($\modS_{\overline{\brac{\ell,\lambda}} \overline{\brac{\ell',\lambda'}}}$).  Quite aside from the point that we would have no use for such entries, it is important to emphasise that the above treatment of atypical irreducible characters makes it clear that we are choosing the characters of the irreducible typicals $\sfmod{\ell}{\TypMod{\lambda}}$, for $\lambda \neq \pm \tfrac{1}{2}$, and the \emph{indecomposable} atypicals $\sfmod{\ell}{\TypMod{\pm 1/2}^+}$ as our basis.  The complete set of S-matrix entries was therefore given in \secref{sec:ModTyp}.  \eqnref{eqn:AtypSMat} is just a dependent quantity that will be useful for studying atypical irreducibles in what follows.

\subsection{The Verlinde Formula}

We now turn to a verification of the Verlinde formula.  Because of the continuous spectrum, this will involve an integral much as the S-transformation formulae \eqref{eqn:TypS} and \eqref{eqn:AtypS} do.  We begin by checking the typical fusion rule
\begin{equation} \label{eqn:TypFus}
\sfmod{\ell}{\TypMod{\lambda}} \fuse \sfmod{m}{\TypMod{\mu}} = \sfmod{\ell + m + 1}{\TypMod{\lambda + \mu + 1/2}} \oplus \sfmod{\ell + m - 1}{\TypMod{\lambda + \mu - 1/2}}.
\end{equation}
The Verlinde formula gives
\begin{align}
\fuscoeff{\brac{\ell,\lambda} \brac{m,\mu}}{\brac{n,\nu}} &= \sum_{\ell' \in \ZZ} \int_{-1}^1 \frac{\modS_{\brac{\ell,\lambda} \brac{\ell',\lambda'}} \modS_{\brac{m,\mu} \brac{\ell',\lambda'}} \modS_{\brac{n,\nu} \brac{\ell',\lambda'}}^*}{\modS_{\overline{\brac{0,0}} \brac{\ell',\lambda'}}} \notag \\
&= \frac{1}{2} \sum_{\ell' \in \ZZ} \ee^{\ii \pi \brac{\brac{\ell + m - n} \ell' / 2 - \ell' \brac{\lambda + \mu - \nu}}} \int_{-1}^1 \ee^{-\ii \pi \brac{\ell + m - n} \lambda'} \func{\cos}{\pi \lambda'} \dd \lambda' \notag \\
&= \frac{1}{2} \sum_{\ell' \in \ZZ} \ee^{\ii \pi \brac{\brac{\ell + m - n} \ell' / 2 - \ell' \brac{\lambda + \mu - \nu}}} \brac{\delta_{\ell + m - n = 1} + \delta_{\ell + m - n = -1}} \notag \\
&= \delta_{n = \ell + m + 1} \func{\delta}{\nu = \lambda + \mu + \tfrac{1}{2} \bmod 2} + \delta_{n = \ell + m - 1} \func{\delta}{\nu = \lambda + \mu - \tfrac{1}{2} \bmod 2}.
\end{align}
This is in perfect agreement with \eqref{eqn:TypFus} because it predicts the Grothendieck ``fusion rule''
\begin{align}
\ch{\sfmod{\ell}{\TypMod{\lambda}}} \fuse \ch{\sfmod{m}{\TypMod{\mu}}} &= \sum_{n \in \ZZ} \int_{-1}^1 \fuscoeff{\brac{\ell,\lambda} \brac{m,\mu}}{\brac{n,\nu}} \ch{\sfmod{n}{\TypMod{\nu}}} \dd \nu \notag \\
&= \ch{\sfmod{\ell + m + 1}{\TypMod{\lambda + \mu + 1/2}}} + \ch{\sfmod{\ell + m - 1}{\TypMod{\lambda + \mu - 1/2}}}.
\end{align}
We remark that this confirmation of the Verlinde formula gives strong additional support to the conjecture \eqref{eqn:FusionAssumption} concerning the interplay between fusion and spectral flow.

The Grothendieck fusion rule corresponding to $\sfmod{\ell}{\IrrMod{\lambda}} \fuse \sfmod{m}{\TypMod{\mu}} = \sfmod{\ell + m}{\TypMod{\lambda + \mu}}$ follows from the Verlinde formula even more easily.  The resulting fusion coefficients are
\begin{equation}
\fuscoeff{\overline{\brac{\ell,\lambda}} \brac{m,\mu}}{\brac{n,\nu}} = \delta_{n = \ell + m} \func{\delta}{\nu = \lambda + \mu \bmod 2},
\end{equation}
as expected.  The derivation corresponding to $\sfmod{\ell}{\IrrMod{\lambda}} \fuse \sfmod{m}{\IrrMod{\mu}} = \sfmod{\ell + m}{\IrrMod{\lambda + \mu}}$ is slightly less straight-forward and deserves comment.  Indeed, one computes that
\begin{subequations}
\begin{align}
&\fuscoeff{\overline{\brac{\ell,\lambda}} \overline{\brac{m,\mu}}}{\brac{n,\nu}} = \frac{1}{8} \sum_{\ell' \in \ZZ} \ee^{\ii \pi \sqbrac{\brac{\ell + m - n} / 2 - \brac{\lambda + \mu - \nu}} \ell'} \int_{-1}^1 \frac{\ee^{-\ii \pi \brac{\ell + m - n} \lambda'}}{\func{\cos}{\pi \lambda'}} \dd \lambda' \label{eqn:Divergent} \\
&\mspace{50mu} = \frac{1}{2} \func{\delta}{\nu = \lambda + \mu - \tfrac{1}{2} \brac{\ell + m -n} \bmod 2} \sum_{\ell'' = 0}^{\infty} \brac{-1}^{\ell''} \int_{-1}^1 \ee^{-\ii \pi \brac{\ell + m - n + 1 + 2 \ell''} \lambda'} \dd \lambda' \notag \\
&\mspace{50mu} = 
\begin{cases}
\brac{-1}^{\brac{n - \ell - m - 1}/2} \func{\delta}{\nu = \lambda + \mu - \tfrac{1}{2} \brac{\ell + m -n} \bmod 2}, & \text{if $n - \ell - m - 1 \in 2 \NN$,} \\
0, & \text{otherwise.}
\end{cases}
\end{align}
\end{subequations}
The alert reader will have noticed that the integral in \eqref{eqn:Divergent} is actually divergent.  We recall that the cosine function appearing here, and in the denominators of \eqref{eqn:AtypSMat} and \eqref{eqn:VacSMat}, results from summing the geometric series appearing in \eqref{eqn:GeomSeries} at the boundary of its radius of convergence.  These denominators should therefore be understood as formal placeholders for the geometric series.  The above computation proceeds smoothly once this series has been correctly reinstated in \eqref{eqn:Divergent} and it is then easy to check that
\begin{equation}
\ch{\sfmod{\ell}{\IrrMod{\lambda}}} \fuse \ch{\sfmod{m}{\IrrMod{\mu}}} = \sum_{\ell' = 0}^{\infty} \brac{-1}^{\ell'} \ch{\sfmod{\ell + m + 2 \ell' + 1}{\TypMod{\lambda + \mu + \ell' + 1/2}}} = \ch{\sfmod{\ell + m}{\IrrMod{\lambda + \mu}}}.
\end{equation}

This shows that the Verlinde formula correctly reproduces the Grothendieck fusion rules involving the irreducible typicals $\sfmod{\ell}{\TypMod{\lambda}}$ and the irreducible atypicals $\sfmod{\ell}{\IrrMod{\lambda}}$.  It is routine to check that this is also consistent with the fusion rules of the atypical indecomposables $\sfmod{\ell}{\StagMod{\lambda}}$; such checks are therefore omitted.  We remark only that we will not use the Verlinde formula to compute fusion coefficients of the form $\fuscoeff{ab}{c}$ in which $c$ refers to an atypical irreducible (and do not expect that such an application would be meaningful).  The reason is again that the atypical irreducibles do not belong to the basis of characters which we have chosen to work with.

\section{Modular Invariants and Extended Algebras} \label{sec:ModInv}

\subsection{Modular Invariants}

With the modular properties of the characters well in hand, we may now turn to the question of constructing modular invariant partition functions.  The most obvious candidate is the diagonal partition function, which in our context is given by
\begin{equation} \label{MI:diag}
\partfunc_{\text{diag.}} = \sum_{\ell \in \ZZ} \int_{-1}^1 \overline{\ch{\sfmod{\ell}{\TypMod{\lambda}}}} \ch{\sfmod{\ell}{\TypMod{\lambda}}} \dd \lambda.
\end{equation}
T-invariance is obvious and S-invariance follows directly from unitarity (\eqnref{eqn:SProps}).  The invariance of the charge-conjugation partition function
\begin{equation} \label{MI:cc}
\partfunc_{\text{c.c.}} = \sum_{\ell \in \ZZ} \int_{-1}^1 \overline{\ch{\sfmod{\ell}{\TypMod{\lambda}}}} \ch{\sfmod{-\ell}{\TypMod{-\lambda}}} \dd \lambda.
\end{equation}
is likewise deduced from \eqnDref{eqn:TypT}{eqn:SProps}.  We remark that the atypical regime $\lambda = \pm \tfrac{1}{2}$ cannot be ignored in these invariants, despite representing a set of measure zero, because it is where the vacuum of the theory resides.  Indeed, the character relations
\begin{equation}
\ch{\sfmod{\ell}{\TypMod{\lambda}^{\pm}}} = \ch{\sfmod{\ell - 1}{\IrrMod{\lambda - 1/2}}} + \ch{\sfmod{\ell + 1}{\IrrMod{\lambda + 1/2}}}, \qquad \text{($\lambda = \pm \tfrac{1}{2}$),}
\end{equation}
show that the combination $\overline{\ch{\IrrMod{0}}} \ch{\IrrMod{0}}$ occurs with multiplicity two in the integrands of both $\partfunc_{\text{diag.}}$ and $\partfunc_{\text{c.c.}}$.  This does not imply that these theories have two vacua (or indeed, \emph{any} vacua) because the partition function does not tell us if such states are eigenvectors, or merely generalised eigenvectors, of $L_0$.  Nevertheless, $\abs{\ch{\IrrMod{0}}}^2$ occurring is necessary for the corresponding theory to have a vacuum.

One can, in fact, construct infinitely many different modular invariants from the admissible spectrum.  Classifying these is an interesting problem, but one that is well beyond our present aims.  Instead, we will consider certain invariants built from a \emph{discrete} subset of the spectrum.  In particular, we recall \cite{RidSL210} that the coset relation between $\AKMA{sl}{2}_{-1/2}$ and the triplet algebra $W \brac{1,2}$ strongly suggests that there should exist a modular invariant of the former involving the atypical modules and the typical irreducibles $\sfmod{\ell}{\TypMod{0}}$ and $\sfmod{\ell}{\TypMod{1}}$.  Moreover, the $\sfmod{\ell}{\TypMod{0}}$ should not couple to the $\sfmod{\ell'}{\TypMod{1}}$ (because their coset versions do not in the coset modular invariant).

We begin by analysing the constraints of T-invariance for this special case.  If $\sfmod{\ell}{\TypMod{\lambda}}$ and $\sfmod{m}{\TypMod{\mu}}$ are to be coupled in a T-invariant partition function, then \eqnref{eqn:TypT} requires that
\begin{equation}
\ell \brac{\lambda - \frac{\ell}{4}} = m \brac{\mu - \frac{m}{4}} \bmod{2}.
\end{equation}
When $\lambda = \mu$, this simplifies rather nicely as
\begin{equation}
\brac{\ell - 2 \lambda}^2 = \brac{m - 2 \lambda}^2 \bmod{8},
\end{equation}
from which we obtain (for $2 \lambda \in \ZZ$) that $\ell$ and $m$ either have the opposite parity to $2 \lambda$, or they have the same parity as $2 \lambda$ and $\ell - m$ is divisible by $4$.  The simplest condition guaranteeing T-invariance, when $\lambda \in \tfrac{1}{2} \ZZ$, is therefore to take $\ell = m \bmod{4}$.

Based on this observation, we propose the following (T-invariant) partition function:
\begin{equation}
\partfunc_2 = \underset{\ell = m \bmod{4}}{\sum_{\ell \in \ZZ} \sum_{m \in \ZZ}} \sum_{j=0}^3 \overline{\ch{\sfmod{\ell}{\TypMod{j/2}}}} \ch{\sfmod{m}{\TypMod{j/2}}}.
\end{equation}
As with the diagonal and charge-conjugation invariants above, $\abs{\ch{\IrrMod{0}}}^2$ appears with multiplicity $2$, so the vacuum can be accommodated within this proposal.  It only remains to check S-invariance.  For this, one applies \eqref{eqn:TypS}, obtaining two more sums over spectral flow parameters $\ell'$ and $m'$ and two integrals over $\SLA{sl}{2}$-weights $\lambda'$ and $\mu'$.  The phases from the S-matrix elements factorise in such a way that the sums over $\ell$, $m = \ell + 4n$ and $j$ may be performed.  The first two give
\begin{equation}
2 \func{\delta}{\lambda' = \tfrac{1}{2} \ell' + \mu' - \tfrac{1}{2} m' \bmod{2}} \qquad \text{and} \qquad \frac{1}{2} \func{\delta}{\mu' = \tfrac{1}{2} m' \bmod{\tfrac{1}{2}}},
\end{equation}
respectively, while the third is a geometric series which sums to $0$ unless $\ell' - m'$ is divisible by $4$, in which case the sum is $4$.  Noting the ``$\bmod{\tfrac{1}{2}}$'' in the second delta function's argument, we therefore obtain
\begin{equation}
\modS \set{\partfunc_2} = \underset{\ell' = m' \bmod{4}}{\sum_{\ell' \in \ZZ} \sum_{m' \in \ZZ}} \sum_{j'=0}^3 \overline{\ch{\sfmod{\ell'}{\TypMod{\brac{\ell' + j'} / 2}}}} \ch{\sfmod{m'}{\TypMod{\brac{m' + j'} / 2}}} = \partfunc_2,
\end{equation}
because shifting $j'$ to $j = \ell' + j'$ shifts $m' + j'$ to $m' - \ell' + j = j \bmod{4}$.  We contend that $\partfunc_2$ is the partition function of the $\AKMA{sl}{2}_{-1/2}$-theory whose coset is the triplet model $W \brac{1,2}$.

In fact, $\partfunc_2$ is the first of an infinite sequence of discrete modular invariant partition functions.  Take $b$ to be a positive even integer.  Then, we claim that
\begin{equation} \label{eqn:Zb}
\partfunc_b = \underset{\ell = m \bmod{2b}}{\sum_{\ell \in \ZZ} \sum_{m \in \ZZ}} \sum_{j=0}^{2b-1} \overline{\ch{\sfmod{\ell}{\TypMod{j/b}}}} \ch{\sfmod{m}{\TypMod{j/b}}}
\end{equation}
is modular invariant.  As $b$ is even, we see that the terms with $j = \tfrac{1}{2} b$ and $j = \tfrac{3}{2} b$ will contribute atypical characters in which $\abs{\ch{\IrrMod{0}}}^2$ appears twice.  T-invariance follows from
\begin{equation}
\brac{\ell - \frac{2j}{b}}^2 - \brac{m - \frac{2j}{b}}^2 = \brac{\ell - m} \brac{\ell + m} - 8j \frac{\ell - m}{2b} = 0 \bmod{8},
\end{equation}
which itself follows from $\ell - m \in 2b \ZZ \subseteq 4 \ZZ$ and $\ell + m \in 2 \ZZ$.  Finally, S-invariance follows exactly as with $b=2$, the parity of $b$ only being used to shift $j'$ in the final step.

\subsection{Extended Algebras}

One can interpret the modular invariants $\partfunc_b$ in terms of simple current extensions of the algebra $\AKMA{sl}{2}_{-1/2}$.  Assuming \eqnref{eqn:FusionAssumption}, the twisted vacuum modules $\sfmod{\ell}{\IrrMod{0}}$ are all simple currents of infinite order.  Their states will have integer dimensions precisely when $\ell$ is a multiple of $4$.  For $b$ an even integer, we may construct an extended algebra $\WAlg{b}$ as in \cite{RidSU206,RidMin07} using the fields associated with $\sfmod{2b}{\IrrMod{0}}$.  As an $\AKMA{sl}{2}_{-1/2}$-module, the algebra decomposes into
\begin{equation}
\WAlg{b} \cong \bigoplus_{\ell \in \ZZ} \sfmod{2b \ell}{\IrrMod{0}}, \qquad \text{($b \in 2 \ZZ_+$).}
\end{equation}
Restricting the simple current so that its states have integer dimensions guarantees that the character of this decomposition will be an eigenvector for the modular T-transformation.  One can also construct modules for the extended algebra by combining $\AKMA{sl}{2}_{-1/2}$-modules along orbits under fusion with the chosen simple current.  In particular, we consider the $\WAlg{b}$-modules $\WTypMod{\lambda}{b}$ which decompose into $\AKMA{sl}{2}_{-1/2}$-modules as
\begin{equation}
\WTypMod{\lambda}{b} \cong \bigoplus_{\ell \in \ZZ} \sfmod{2b \ell}{\TypMod{\lambda}}, \qquad \text{($b \in 2 \ZZ_+$).}
\end{equation}
Spectral flow then generates, for each $\lambda$, an additional $2b-1$ non-isomorphic $\WAlg{b}$-modules.  It is easy to check that the corresponding characters will be T-eigenvectors, hence that the extended algebra generators will act with \emph{integer-moding}, if and only if $b \lambda \in \ZZ$.  The natural diagonal partition function for a $\WAlg{b}$-theory is then
\begin{equation}
\partfunc_{\text{diag.}} \bigl[ \WAlg{b} \bigr] = \sum_{r=0}^{2b-1} \sum_{s=0}^{2b-1} \overline{\ch{\sfmod{r}{\WTypMod{s/b}{b}}}} \ch{\sfmod{r}{\WTypMod{s/b}{b}}}, \qquad \text{($b \in 2 \ZZ_+$).}
\end{equation}
Decomposing this combination into $\AKMA{sl}{2}_{-1/2}$-characters, we identify $\partfunc \bigl[ \WAlg{b} \bigr]$ as the invariant partition function $\partfunc_b$.

There are other simple currents with which we can extend $\AKMA{sl}{2}_{-1/2}$.  In particular, the twisted irreducibles $\sfmod{\ell}{\IrrMod{1}}$ are simple currents of integer dimension for all $\ell = 2 \bmod{4}$.  Indeed, $\sfmod{2}{\IrrMod{1}}$ generates the (cyclic subgroup of) integer dimension simple currents.  With these, one can construct the extended algebras
\begin{equation}
\WAlg{b} \cong \bigoplus_{\ell \in \ZZ} \sqbrac{\sfmod{4b \ell}{\IrrMod{0}} \oplus \sfmod{2b \brac{2 \ell + 1}}{\IrrMod{1}}}, \qquad \text{($b \in 2 \ZZ_+ - 1$),}
\end{equation}
whose (untwisted) $\TypMod{}{}$-type modules are given by
\begin{equation}
\WTypMod{\lambda}{b} \cong \bigoplus_{\ell \in \ZZ} \sqbrac{\sfmod{4b \ell}{\TypMod{\lambda}} \oplus \sfmod{2b \brac{2 \ell + 1}}{\TypMod{\lambda + 1}}} \cong \WTypMod{\lambda}{2b} \oplus \sfmod{2b}{\WTypMod{\lambda + 1}{2b}}, \qquad \text{($b \in 2 \ZZ_+ - 1$),}
\end{equation}
when restricted appropriately.  Noting that $\sfmod{2b}{\WTypMod{\lambda}{b}} = \WTypMod{\lambda + 1}{b}$ and that T-invariance requires $b \lambda \in \ZZ + \tfrac{1}{2}$, the diagonal partition function would have the form
\begin{align}
\partfunc_{\text{diag.}} \bigl[ \WAlg{b} \bigr] &= \sum_{r=0}^{2b-1} \sum_{s=0}^{2b-1} \overline{\ch{\sfmod{r}{\WTypMod{s/b-1/2}{b}}}} \ch{\sfmod{r}{\WTypMod{s/b-1/2}{b}}} \qquad \text{($b \in 2 \ZZ_+ - 1$)} \notag \\
&= \sum_{r=0}^{4b-1} \sum_{s=0}^{2b-1} \sqbrac{\abs{\ch{\sfmod{r}{\WTypMod{s/b-1/2}{b}}}}^2 + \overline{\ch{\sfmod{r}{\WTypMod{s/b-1/2}{b}}}} \ch{\sfmod{r+2}{\WTypMod{s/b+1/2}{b}}}}.
\end{align}
However, one can check that this expression is not S-invariant.  While there are undoubtedly other modular invariants to find, they are not constructed from the simple currents $\sfmod{\ell}{\IrrMod{1}}$.\footnote{We also mention that one can extend by the simple current $\sfmod{2b}{\IrrMod{0}}$ with $b$ odd.  The dimensions of the extension fields are now \emph{half-integers}, so one expects fermionic behaviour.  However, we expect that the modular invariant constructed using supercharacters and half-integer moded sectors will coincide with the bosonic invariant $\partfunc_{2b}$.}

The reason for this is ultimately rooted in the fact that the simple currents $\sfmod{\ell}{\IrrMod{1}}$, unlike the $\sfmod{\ell}{\IrrMod{0}}$, cannot be obtained from the vacuum module by twisting with an automorphism of $\AKMA{sl}{2}_{-1/2}$.  More concretely, observe that the spectral flow automorphism $\sfaut^{2b}$ satisfies
\begin{equation}
\modS_{\sfaut^{2b} \brac{\ell,\lambda} \brac{\ell',\lambda'}} = \modS_{\brac{\ell + 2b,\lambda} \brac{\ell',\lambda'}} = \ee^{\ii \pi b \ell'} \ee^{-2 \pi \ii b \lambda'} \modS_{\brac{\ell,\lambda} \brac{\ell',\lambda'}},
\end{equation}
so that the S-matrix is left unchanged when $b$ is even and $b \lambda' \in \ZZ$.  The transformation rule for $\omega \colon \brac{\ell,\lambda} \mapsto \brac{\ell + 2b,\lambda + 1}$ is rather
\begin{equation}
\modS_{\omega \brac{\ell,\lambda} \brac{\ell',\lambda'}} = \ee^{\ii \pi \brac{b-1} \ell'} \ee^{-2 \pi \ii b \lambda'} \modS_{\brac{\ell,\lambda} \brac{\ell',\lambda'}}.
\end{equation}
For $b$ odd, one can therefore guarantee S-invariance ($b \lambda' \in \ZZ$) or T-invariance ($b \lambda' \in \ZZ + \tfrac{1}{2}$), but not both.

Returning to $b$ even, we remark that the extended algebra theories with partition functions $\partfunc \bigl[ \WAlg{b} \bigr] = \partfunc_b$ are \emph{rational} in the sense that they are constructed from a finite number of irreducible representations.  However, their modular behaviour is not exemplary.  It is straight-forward to compute the S-matrix for the characters of the $\sfmod{\ell}{\WTypMod{\lambda}{b}}$ using \eqref{eqn:TypSMat}:
\begin{equation}
\WmodS_{\brac{\ell,\lambda} \brac{\ell',\lambda'}} = \frac{1}{2b} \frac{\abs{\tau}}{-\ii \tau} \ee^{\ii \pi \brac{\ell \ell'/2 - \ell \lambda' - \ell' \lambda}}.
\end{equation}
This is a $4b^2 \times 4b^2$ matrix, symmetric and unitary.  However, one quickly runs into difficulty when attempting to apply this to a Verlinde formula:  The trick using resolutions to construct the atypical irreducible characters, and the vacuum character in particular, can no longer be used because the periodicity $\sfmod{2b}{\WTypMod{\lambda}{b}} = \WTypMod{\lambda}{b}$ leads to divergences.  Summing the S-matrix contributions \eqref{eqn:AtypSMat} directly likewise leads to divergences --- the matrix elements $\WmodS_{\overline{\brac{\ell,\lambda}} \brac{\ell',\lambda'}}$ are undefined when $\lambda'$ takes on an atypical value.  This bad modular behaviour seems to be a fundamental problem with extended theories.  Indeed, it is straight-forward to deduce the (Grothendieck) fusion rules of the $\WAlg{b}$-theories and see that the fusion matrices are not diagonalisable, hence that the standard Verlinde formula cannot apply.  In this, the extended theories are similar to the $W \brac{1,p}$-models studied, for example, in \cite{FloMod96,FucNon04,FeiMod06,GaiRad09,PeaGro10,TsuTen12}.  We expect that a detailed investigation of the extended theories constructed here will help to shed further light on the modular properties of rational \lcfts{} in general.

\section{Models with $\AKMA{sl}{2}_{-4/3}$-Symmetry} \label{sec:k=-4/3}

In this section, we summarise the result of applying the methods described above for $k=-\tfrac{1}{2}$ to the case when the level is $k=-\tfrac{4}{3}$ and the central charge is $c=-6$.  This is the only other level for which the fusion rules and resulting indecomposable structures have been (partially) computed and analysed, respectively \cite{GabFus01,AdaLat09}.  We include this case to demonstrate that our approach to modular transformations and the Verlinde formula is quite robust.  Indeed, we shall use it to complete the structural analysis of the staggered modules generated by fusion and to provide a non-trivial check of a conjectured fusion rule.  The generalisation of the above formalism to arbitrary admissible level $\AKMA{sl}{2}$-theories, for which there is no detailed description of the fusion rules, will be discussed elsewhere.

\subsection{Spectrum and Fusion}

The admissible $k=-\tfrac{4}{3}$ irreducibles from our relaxed category once again fall into two countably-infinite families and one uncountably-infinite family:
\begin{equation*}
\sfmod{\ell}{\IrrMod{0}}, \quad \sfmod{\ell}{\DiscMod{-2/3}^+}, \quad \sfmod{\ell}{\TypMod{\lambda}} \qquad \text{($\ell \in \ZZ$, $\lambda \in \left( -1 , 1 \right] \setminus \set{\pm \tfrac{2}{3}}$).}
\end{equation*}
Here, the minimal conformal dimension among the states of $\TypMod{\lambda}$ is $-\tfrac{1}{3}$ and we shall use the following identifications liberally in the rest of this section:
\begin{equation}
\sfmod{}{\IrrMod{0}} = \DiscMod{-4/3}^+, \qquad \sfmod{-1}{\IrrMod{0}} = \DiscMod{4/3}^-, \qquad \sfmod{-1}{\DiscMod{-2/3}^+} = \DiscMod{2/3}^-.
\end{equation}
This spectrum is illustrated for convenience in \figref{fig:Spec2}.

{
\psfrag{L0}[][]{$\IrrMod{0}$}
\psfrag{L1}[][]{$\IrrMod{1}$}
\psfrag{E0}[][]{$\TypMod{\lambda}$}
\psfrag{La}[][]{$\DiscMod{-4/3}^+$}
\psfrag{Da}[][]{$\DiscMod{-2/3}^+$}
\psfrag{La*}[][]{$\DiscMod{4/3}^-$}
\psfrag{Da*}[][]{$\DiscMod{2/3}^-$}
\psfrag{g}[][]{$\sfaut$}
\psfrag{00}[][]{$\scriptstyle \brac{0,0}$}
\psfrag{aa}[][]{$\scriptstyle \tbrac{-\tfrac{4}{3},-\tfrac{1}{3}}$}
\psfrag{bb}[][]{$\scriptstyle \tbrac{\tfrac{4}{3},-\tfrac{1}{3}}$}
\psfrag{cc}[][]{$\scriptstyle \tbrac{-\tfrac{8}{3},-\tfrac{4}{3}}$}
\psfrag{dd}[][]{$\scriptstyle \tbrac{\tfrac{8}{3},-\tfrac{4}{3}}$}
\psfrag{ee}[][]{$\scriptstyle \tbrac{-\tfrac{2}{3},-\tfrac{1}{3}}$}
\psfrag{ff}[][]{$\scriptstyle \tbrac{\tfrac{2}{3},-\tfrac{1}{3}}$}
\psfrag{gg}[][]{$\scriptstyle \brac{-2,-1}$}
\psfrag{hh}[][]{$\scriptstyle \brac{2,-1}$}
\psfrag{0e}[][]{$\scriptstyle \tbrac{\lambda,-\tfrac{1}{3}}$}
\psfrag{aq}[][]{$\scriptstyle \tbrac{\lambda-\tfrac{4}{3},\tfrac{\lambda}{2}-\tfrac{2}{3}}$}
\psfrag{cq}[][]{$\scriptstyle \tbrac{\lambda-\tfrac{8}{3},\lambda-\tfrac{5}{3}}$}
\psfrag{bq}[][]{$\scriptstyle \tbrac{\lambda+\tfrac{4}{3},-\tfrac{\lambda}{2}-\tfrac{2}{3}}$}
\psfrag{dq}[][]{$\scriptstyle \tbrac{\lambda+\tfrac{8}{3},-\lambda-\tfrac{5}{3}}$}
\begin{figure}
\begin{center}
\includegraphics[width=\textwidth]{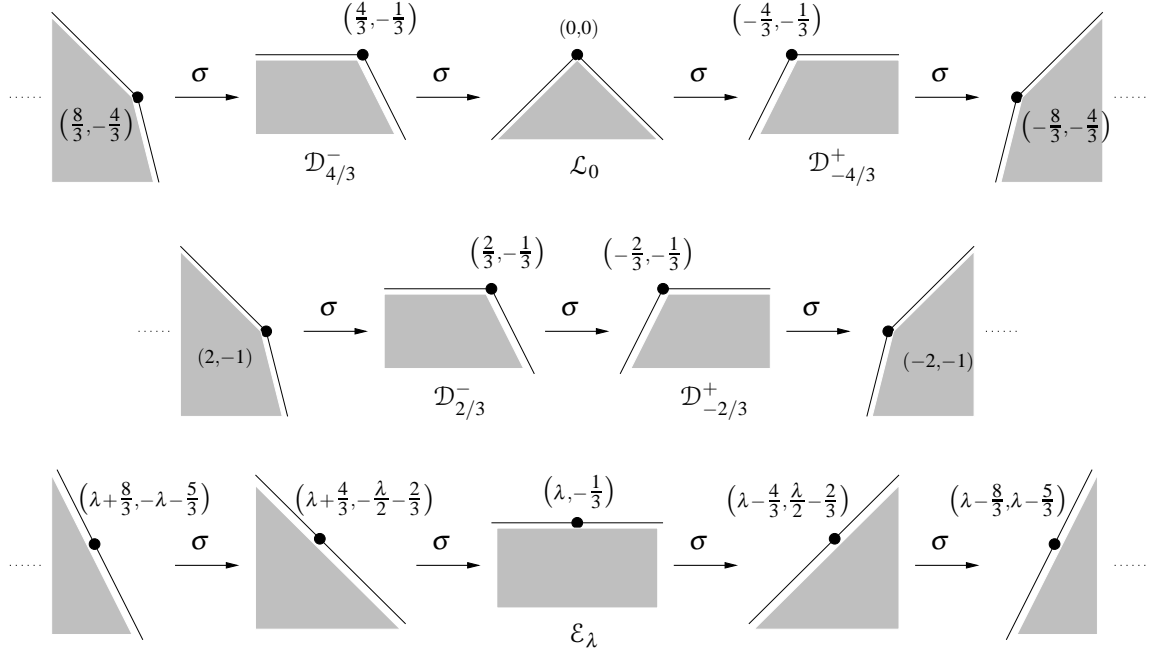}
\caption{Depictions of the admissible irreducible $\AKMA{sl}{2}_{-4/3}$-modules.  Each labelled state declares its $\func{\alg{sl}}{2}$-weight and conformal dimension (in that order).  Conformal dimensions increase from top to bottom and $\SLA{sl}{2}$-weights increase from right to left.  We have shifted the middle row to emphasise the conjugation symmetry.  Moreover, the modules of this row do behave in many respects as if they should be assigned a half-integer spectral flow index.} \label{fig:Spec2}
\end{center}
\end{figure}
}

Regarding the $\TypMod{\lambda}$ with $\lambda \neq \pm \tfrac{2}{3}$ as typical, we have four atypical $\TypMod{}$-type modules whose structure diagrams are
\begin{center}
\parbox[c]{0.85\textwidth}{
\begin{tikzpicture}[thick,
	nom/.style={circle,draw=black!20,fill=black!20,inner sep=1pt}
	]
\node (t0) at (1,2) [] {$\DiscMod{2/3}^-$};
\node (l0) at (0,1) [nom] {$\TypMod{+2/3}^+$};
\node (b0) at (1,0) [] {$\DiscMod{-4/3}^+$};
\draw [->] (t0) -- (b0);
\node (t1) at (4.5,2) [] {$\DiscMod{-2/3}^+$};
\node (l1) at (3.5,1) [nom] {$\TypMod{-2/3}^-$};
\node (b1) at (4.5,0) [] {$\DiscMod{4/3}^-$};
\draw [->] (t1) -- (b1);
\node (t2) at (8,2) [] {$\DiscMod{4/3}^-$};
\node (l2) at (7,1) [nom] {$\TypMod{-2/3}^+$};
\node (b2) at (8,0) [] {$\DiscMod{-2/3}^+$};
\draw [->] (t2) -- (b2);
\node (t3) at (11.5,2) [] {$\DiscMod{-4/3}^+$};
\node (l3) at (10.5,1) [nom] {$\TypMod{+2/3}^-$};
\node (b3) at (11.5,0) [] {$\DiscMod{2/3}^-$};
\draw [->] (t3) -- (b3);
\end{tikzpicture}
}.
\end{center}
As with the case $k=-\tfrac{1}{2}$, further indecomposables may be constructed through fusion.  Assuming \eqref{eqn:FusionAssumption}, we can summarise the computations of \cite{GabFus01} as follows:
\begin{equation} \label{FR:Gab}
\DiscMod{2/3}^- \fuse \DiscMod{-2/3}^+ = \IrrMod{0} \oplus \TypMod{0}, \qquad \DiscMod{-2/3}^+ \fuse \TypMod{0} = \StagMod{-2/3}^+, \qquad \TypMod{0} \fuse \TypMod{0} = \TypMod{0} \oplus \StagMod{0}.
\end{equation}
Of course, $\IrrMod{0}$ acts as the fusion identity.  Both $\StagMod{0}$ and $\StagMod{-2/3}^+$ are indecomposable modules\footnote{We find it natural to affix a superscript ``$+$'' to $\StagMod{-2/3}^+$ as this module is not expected to be self-conjugate.  Instead, we expect that its conjugate is $\conjmod{\StagMod{-2/3}^+} = \sfmod{-1}{\StagMod{-2/3}^+}$, which we will therefore denote by $\StagMod{2/3}^-$.} on which $L_0$ acts non-diagonalisably, and the same is true for their images under spectral flow.  We therefore regard them as staggered modules.  Unfortunately, while the analysis in \cite{GabFus01} provides detailed structural information for the first two filtered quotients employed by the fusion algorithm, it was not clear if this information is sufficient to derive full structure diagrams and characters.  We will confirm in what follows that this is indeed the case.

The fusion rules involving the other typical $\TypMod{\lambda}$ were not considered in \cite{GabFus01}.  However, the fusion rules involving the staggered modules $\StagMod{0}$ and $\StagMod{-2/3}^+$ are noted to follow from associativity and the following conjectured rule:
\begin{equation} \label{FR:GabConj}
\DiscMod{-2/3}^+ \fuse \StagMod{0} = \StagMod{-2/3}^+.
\end{equation}
This conjecture was again supported by explicitly constructing certain filtered quotients of the fusion product.  For completeness, we remark that this conjecture implies that
\begin{equation} \label{FR:GabAss}
\begin{aligned}
\TypMod{0} \fuse \StagMod{0} &= 2 \TypMod{0}, \\
\DiscMod{2/3}^- \fuse \StagMod{-2/3}^+ &= 2 \TypMod{0} \oplus \StagMod{0},
\end{aligned}
\qquad
\begin{aligned}
\StagMod{0} \fuse \StagMod{0} &= 2 \StagMod{0}, \\
\TypMod{0} \fuse \StagMod{-2/3}^+ &= 2 \StagMod{-2/3}^+,
\end{aligned}
\qquad
\begin{aligned}
\StagMod{0} \fuse \StagMod{-2/3}^+ &= 2 \StagMod{-2/3}^+, \\
\StagMod{2/3}^- \fuse \StagMod{-2/3}^+ &= 4 \TypMod{0} \oplus 2 \StagMod{0}.
\end{aligned}
\end{equation}
\eqnref{eqn:FusionAssumption} is assumed to extend these rules to their spectrally-flowed versions.

\subsection{Characters} \label{sec:Char-4/3}

The characters of the admissible highest weight modules are readily determined:
\begin{equation} \label{ch:IHW}
\begin{split}
\ch{\IrrMod{0}} &= \frac{y^{-4/3} q^{1/4} \sum_{n \in \ZZ} \brac{-1}^n z^{2n} q^{3n \brac{n+1} / 2}}{\prod_{i=1}^{\infty} \brac{1 - z^2 q^i} \brac{1-q^i} \brac{1 - z^{-2} q^{i-1}}} = y^{-4/3} \frac{\Jth{1}{z^2 ; q^3}}{\Jth{1}{z^2 ; q}}, \\
\ch{\DiscMod{-4/3}^+} &= \frac{y^{-4/3} z^{-4/3} q^{-1/12} \sum_{n \in \ZZ} \brac{-1}^n z^{2n} q^{n \brac{3n-1} / 2}}{\prod_{i=1}^{\infty} \brac{1 - z^2 q^i} \brac{1-q^i} \brac{1 - z^{-2} q^{i-1}}} = y^{-4/3} z^{-4/3} q^{2/3} \frac{\Jth{1}{z^2 q^{-2} ; q^3}}{\Jth{1}{z^2 ; q}}, \\
\ch{\DiscMod{-2/3}^+} &= \frac{y^{-4/3} z^{-2/3} q^{-1/12} \sum_{n \in \ZZ} \brac{-1}^n z^{2n} q^{n \brac{3n+1} / 2}}{\prod_{i=1}^{\infty} \brac{1 - z^2 q^i} \brac{1-q^i} \brac{1 - z^{-2} q^{i-1}}} = y^{-4/3} z^{-2/3} q^{1/6} \frac{\Jth{1}{z^2 q^{-1} ; q^3}}{\Jth{1}{z^2 ; q}}.
\end{split}
\end{equation}
As meromorphic functions, these formulae lead to the relations
\begin{equation} \label{eqn:ChRels}
\ch{\sfmod{\ell-1}{\DiscMod{-2/3}^+}} + \ch{\sfmod{\ell+1}{\IrrMod{0}}} = \ch{\sfmod{\ell-1}{\IrrMod{0}}} + \ch{\sfmod{\ell}{\DiscMod{-2/3}^+}} = 0,
\end{equation}
which demonstrates that the space spanned by the $\ch{\sfmod{\ell}{\IrrMod{0}^+}}$ and the $\ch{\sfmod{\ell}{\DiscMod{-2/3}^+}}$ again has a basis given by the characters of the highest weight admissibles.  Moreover, the atypical indecomposables $\TypMod{\pm 1/2}^{\pm}$ again have vanishing meromorphic characters.

The annulus of convergence in which one should expand the character of $\sfmod{\ell}{\IrrMod{0}}$ is given by
\begin{equation}
\abs{q} < 1, \qquad \abs{q}^{-\brac{\ell-1}} < \abs{z}^2 < \abs{q}^{-\brac{\ell+1}}.
\end{equation}
This contrasts with the annulus for that of $\sfmod{\ell}{\DiscMod{-2/3}^+}$ which is given by
\begin{equation}
\abs{q} < 1, \qquad \abs{q}^{-\ell} < \abs{z}^2 < \abs{q}^{-\brac{\ell+1}}.
\end{equation}
We note that the relations \eqref{eqn:ChRels} again correspond to summing characters whose natural regions of expansion are disjoint.

To compute the characters of the atypicals $\TypMod{\pm 2/3}^{\pm}$, we again use the identity of Kac and Wakimoto detailed in \appref{app:Magic}.  For $\TypMod{2/3}^{\pm}$, we begin by using \eqref{eqn:Magic2} to expand the character of $\DiscMod{-4/3}^+$ in the annulus $1 < \abs{z}^2 < \abs{q}^{-1}$, obtaining
\begin{equation}
\ch{\DiscMod{-4/3}^+} = -\ii y^{-4/3} z^{-4/3} q^{2/3} \frac{\Jth{1}{z^2 q^{-2} ; q^3} \Jth{1}{u ; q}}{\Jth{1}{u z^2 ; q} \func{\eta}{q}^3} \sum_{n \in \ZZ} \frac{z^{2n} u q^n}{1-uq^n}.
\end{equation}
We next compute the character of $\DiscMod{2/3}^-$ and then expand it in the annulus $\abs{q} < \abs{z}^2 < 1$ using \eqref{eqn:Magic1}:
\begin{align}
\ch{\DiscMod{2/3}^-} &= y^{-4/3} z^{2/3} q^{1/6} \frac{\Jth{1}{z^2 q^{-2} ; q^3}}{\Jth{1}{z^2 q^{-1} ; q}} = -y^{-4/3} z^{-4/3} q^{2/3} \frac{\Jth{1}{z^2 q^{-2} ; q^3}}{\Jth{1}{z^2 ; q}} \notag \\
&= \ii y^{-4/3} z^{-4/3} q^{2/3} \frac{\Jth{1}{z^2 q^{-2} ; q^3} \Jth{1}{u ; q}}{\Jth{1}{u z^2 ; q} \func{\eta}{q}^3} \sum_{n \in \ZZ} \frac{z^{2n}}{1-uq^n}.
\end{align}
Adding these expansions (as formal power series) now gives
\begin{align}
\ch{\TypMod{2/3}^{\pm}} &= \ii y^{-4/3} z^{-4/3} q^{2/3} \frac{\Jth{1}{z^2 q^{-2} ; q^3} \Jth{1}{u ; q}}{\Jth{1}{u z^2 ; q} \func{\eta}{q}^3} \sum_{n \in \ZZ} z^{2n} \notag \\
&= \ii y^{-4/3} z^{-4/3} q^{2/3} \sum_{m \in \ZZ} \frac{\Jth{1}{\ee^{2 \pi \ii m} q^{-2} ; q^3} \Jth{1}{u ; q}}{\Jth{1}{\ee^{2 \pi \ii m} u ; q} \func{\eta}{q}^3} \func{\delta}{2 \zeta - m} \notag \\
&= \ii y^{-4/3} z^{2/3} q^{2/3} \frac{\Jth{1}{q^{-2} ; q^3}}{\func{\eta}{q}^3} \sum_{n \in \ZZ} z^{2n} \notag \\
&= y^{-4/3} \frac{z^{2/3}}{\func{\eta}{q}^2} \sum_{n \in \ZZ} z^{2n}. \label{eqn:AtypSum}
\end{align}
A similar calculation (or conjugating) gives an analogous formula for $\ch{\TypMod{-2/3}^{\pm}}$.  We remark that these results agree perfectly with the character formula for typical irreducibles, namely
\begin{equation} \label{ch:Typ-4/3}
\ch{\TypMod{\lambda}} = y^{-4/3} \frac{z^{\lambda}}{\func{\eta}{q}^2} \sum_{n \in \ZZ} z^{2n}.
\end{equation}
Indeed, this can even be derived from \eqref{eqn:AtypSum} using the general structure theory outlined in \cite{FeiEqu98,SemEmb97}.

Finally, consider the fusion ring generated by the modules $\sfmod{\ell}{\IrrMod{0}}$, $\sfmod{\ell}{\DiscMod{-2/3}^+}$, $\sfmod{\ell}{\TypMod{0}}$, $\sfmod{\ell}{\StagMod{0}}$ and $\sfmod{\ell}{\StagMod{-2/3}^+}$ (we exclude the other typical irreducibles as \cite{GabFus01} says nothing about their fusion rules).  The quotient of this ring by the ideal generated by the zero-character modules
\begin{equation}
\sfmod{\ell-1}{\DiscMod{-2/3}^+} \oplus \sfmod{\ell+1}{\IrrMod{0}}, \quad \sfmod{\ell-1}{\IrrMod{0}} \oplus \sfmod{\ell}{\DiscMod{-2/3}^+}, \quad \sfmod{\ell}{\TypMod{0}}, \quad \sfmod{\ell}{\StagMod{0}}, \quad \sfmod{\ell}{\StagMod{-2/3}^+}
\end{equation}
is free of rank $3$.  Taking the equivalence classes $\bigl[ \IrrMod{0} \bigr]$, $\bigl[ \DiscMod{-4/3}^+ \bigr]$ and $\bigl[ \DiscMod{-2/3}^+ \bigr]$ as an ordered basis, the induced product $\chfuse$ on the quotient is given by:

\begin{center}
\setlength{\extrarowheight}{4pt}
\begin{tabular}{C|CCC}
\chfuse & \bigl[ \IrrMod{0} \bigr] & \bigl[ \DiscMod{-4/3}^+ \bigr] & \bigl[ \DiscMod{-2/3}^+ \bigr] \\
\hline
\bigl[ \IrrMod{0} \bigr] & \bigl[ \IrrMod{0} \bigr] & \bigl[ \DiscMod{-4/3}^+ \bigr] & \bigl[ \DiscMod{-2/3}^+ \bigr] \\
\bigl[ \DiscMod{-4/3}^+ \bigr] & \bigl[ \DiscMod{-4/3}^+ \bigr] & -\bigl[ \DiscMod{-2/3}^+ \bigr] & -\bigl[ \IrrMod{0} \bigr] \\
\bigl[ \DiscMod{-2/3}^+ \bigr] & \bigl[ \DiscMod{-2/3}^+ \bigr] & -\bigl[ \IrrMod{0} \bigr] & \bigl[ \DiscMod{-4/3}^+ \bigr]
\end{tabular}
\end{center}

\noindent The reader might like to verify that the Verlinde formula recovers these structure coefficients from the S-matrix obtained \cite{KacMod88} from the meromorphic characters \eqref{ch:IHW}:
\begin{equation}
\chmodS = -\frac{1}{\sqrt{3}} 
\begin{pmatrix}
1 & 1 & -1 \\
1 & \ee^{4 \pi \ii / 3} & -\ee^{2 \pi \ii / 3} \\
-1 & -\ee^{2 \pi \ii / 3} & \ee^{4 \pi \ii / 3}
\end{pmatrix}
.
\end{equation}

\subsection{Modular Transformations}

We begin by applying spectral flow \eqref{eqn:CharSF} to the typical characters \eqref{ch:Typ-4/3}, obtaining
\begin{equation}
\ch{\sfmod{\ell}{\TypMod{\lambda}}} = \frac{\ee^{-8 \pi \ii t / 3} \ee^{2 \pi \ii \ell^2 \tau / 3}}{\func{\eta}{\tau}^2} \sum_{m \in \ZZ} \ee^{\ii \pi \brac{\lambda - 4 \ell / 3} m} \func{\delta}{2 \zeta + \ell \tau - m}.
\end{equation}
Here, as before, we have set $y = \ee^{2 \pi \ii t}$, $z = \ee^{2 \pi \ii \zeta}$ and $q = \ee^{2 \pi \ii \tau}$.  The method detailed in \secref{sec:ModTyp} now easily determines the modular properties of these characters:
\begin{subequations}
\begin{align}
\modS \set{\ch{\sfmod{\ell}{\TypMod{\lambda}}}} &= \sum_{\ell' \in \ZZ} \int_{-1}^1 \modS_{\brac{\ell,\lambda} \brac{\ell',\lambda'}} \ch{\sfmod{\ell'}{\TypMod{\lambda'}}} \dd \lambda', \quad \modS_{\brac{\ell,\lambda} \brac{\ell',\lambda'}} = \frac{1}{2} \frac{\abs{\tau}}{-\ii \tau} \ee^{\ii \pi \brac{4 \ell \ell' / 3 - \ell \lambda' - \ell' \lambda}}, \\
\modT \set{\ch{\sfmod{\ell}{\TypMod{\lambda}}}} &= \ee^{\ii \pi \brac{\ell \brac{\lambda - 2 \ell / 3} - 1/6}} \ch{\sfmod{\ell}{\TypMod{\lambda}}}.
\end{align}
\end{subequations}
Of course, these transformations also apply to the characters of the atypicals $\TypMod{\pm 2/3}^{\pm}$.  We remark that the S-matrix is again symmetric and unitary (\eqnref{eqn:SProps}) and that conjugation is implemented, up to a phase, by $\modS^2$.

To deduce the modular properties of the other atypical characters, we construct resolutions as in \secref{sec:ModAtyp}.  Splicing together the short exact sequences describing the structure diagrams of the $\TypMod{\pm 2/3}^+$, we obtain the resolutions
\begin{equation}
\begin{gathered}
\dres{\IrrMod{0}}{\sfmod{}{\TypMod{-2/3}^+}}{\sfmod{2}{\TypMod{2/3}^+}}{\sfmod{4}{\TypMod{-2/3}^+}}{\sfmod{5}{\TypMod{2/3}^+}}, \\
\dres{\DiscMod{-2/3}^+}{\sfmod{}{\TypMod{2/3}^+}}{\sfmod{3}{\TypMod{-2/3}^+}}{\sfmod{4}{\TypMod{2/3}^+}}{\sfmod{6}{\TypMod{-2/3}^+}}.
\end{gathered}
\end{equation}
Applying spectral flow, these imply the following character identities for the irreducible atypicals:
\begin{equation}
\begin{aligned}
\ch{\sfmod{\ell}{\IrrMod{0}}} &= \sum_{\ell' = 0}^{\infty} \brac{\ch{\sfmod{\ell + 3 \ell' + 1}{\TypMod{-2/3}^+}} - \ch{\sfmod{\ell + 3 \ell' + 2}{\TypMod{2/3}^+}}}, \\
\ch{\sfmod{\ell}{\DiscMod{-2/3}^+}} &= \sum_{\ell' = 0}^{\infty} \brac{\ch{\sfmod{\ell + 3 \ell' + 1}{\TypMod{2/3}^+}} - \ch{\sfmod{\ell + 3 \ell' + 3}{\TypMod{-2/3}^+}}}.
\end{aligned}
\end{equation}
We therefore arrive at the modular transformation properties of the atypical irreducibles by summing over those of the indecomposables.  In particular, we compute that
\begin{equation}
\modS_{\overline{\brac{\ell,0}} \brac{\ell',\lambda'}} = \frac{1}{2} \frac{\abs{\tau}}{-\ii \tau} \frac{\ee^{\ii \pi \ell \brac{4 \ell' / 3 - \lambda'}}}{1 + 2 \func{\cos}{\pi \lambda'}}, \qquad
\modS_{\overline{\brac{\ell,-2/3}} \brac{\ell',\lambda'}} = \frac{\abs{\tau}}{-\ii \tau} \frac{\ee^{\ii \pi \brac{\ell + 1/2} \brac{4 \ell' / 3 - \lambda'}} \func{\cos}{\pi \lambda' / 2}}{1 + 2 \func{\cos}{\pi \lambda'}}.
\end{equation}
As before, the overline indicates that the pair refers to an atypical irreducible ($\sfmod{\ell}{\IrrMod{0}}$ and $\sfmod{\ell}{\DiscMod{-2/3}^+}$, respectively).  We mention that these formulae provide good evidence for the claim that the $\sfmod{\ell}{\DiscMod{-2/3}^+}$ should be assigned a spectral flow index of $\ell + \tfrac{1}{2}$.  Finally, note that we are again using the characters of the typical irreducibles $\sfmod{\ell}{\TypMod{\lambda}}$, supplemented by those of the atypical indecomposables $\sfmod{\ell}{\TypMod{\pm 2/3}^+}$, as the character basis.

\subsection{The Verlinde Formula} \label{sec:Verlinde-4/3}

We first apply the Verlinde formula to the (Grothendieck) fusion of two typicals.  The denominator $1 + \func{\cos}{\pi \lambda'}$ of the vacuum S-matrix entries leads to three contributions:
\begin{align}
\fuscoeff{\brac{\ell,\lambda} \brac{m,\mu}}{\brac{n,\nu}} &= \sum_{\ell' \in \ZZ} \int_{-1}^1 \frac{\modS_{\brac{\ell,\lambda} \brac{\ell',\lambda'}} \modS_{\brac{m,\mu} \brac{\ell',\lambda'}} \modS_{\brac{n,\nu} \brac{\ell',\lambda'}}^*}{\modS_{\overline{\brac{0,0}} \brac{\ell',\lambda'}}} \notag \\
&= \delta_{n = \ell + m - 1} \func{\delta}{\nu = \lambda + \mu - \tfrac{4}{3} \bmod{2}} + \delta_{n = \ell + m} \func{\delta}{\nu = \lambda + \mu \bmod{2}} \notag \\
&\hphantom{\delta_{n = \ell + m - 1} \func{\delta}{\nu = \lambda + \mu - \tfrac{4}{3} \bmod{2}}} + \delta_{n = \ell + m + 1} \func{\delta}{\nu = \lambda + \mu + \tfrac{4}{3} \bmod{2}}.
\end{align}
This is again consistent with the conjectured relation \eqref{eqn:FusionAssumption}.  Setting $\ell = m = 0$ for clarity, the Verlinde formula implies the Grothendieck fusion rule
\begin{equation}
\ch{\TypMod{\lambda}} \fuse \ch{\TypMod{\mu}} = \ch{\sfmod{-1}{\TypMod{\lambda + \mu - 4/3}}} + \ch{\TypMod{\lambda + \mu}} + \ch{\sfmod{}{\TypMod{\lambda + \mu + 4/3}}}.
\end{equation}
Comparing conformal dimensions, we deduce that the fusion of $\TypMod{\lambda}$ and $\TypMod{\mu}$ is completely reducible, except perhaps when $\lambda + \mu = 0, 1, \pm \tfrac{2}{3}$.  When $\lambda + \mu = 1$, we conjecture that the result is also completely reducible.\footnote{This conjecture may be verified using the Nahm-Gaberdiel-Kausch algorithm as in \cite{GabFus01,RidFus10}.  However, a detailed description of this calculation (and other direct fusion computations) is beyond the scope of the present article, and we defer it to a future publication.}  When $\lambda + \mu = 0$, however, \eqnref{FR:Gab} shows that $\sfmod{-1}{\TypMod{2/3}}$ and $\sfmod{}{\TypMod{-2/3}}$ combine (with appropriate superscripts ``$\pm$'') to form the indecomposable $\StagMod{0}$ and that $\TypMod{0}$ splits off as a direct summand.  We therefore deduce the following structure diagram for the staggered module $\StagMod{0}$:
\begin{center}
\parbox[c]{0.3\textwidth}{
\begin{tikzpicture}[thick,
	nom/.style={circle,draw=black!20,fill=black!20,inner sep=1pt}
	]
\node (top) at (0,1.5) [] {$\IrrMod{0}$};
\node (left) at (-1.5,0) [] {$\sfmod{-1}{\DiscMod{2/3}^-}$};
\node (right) at (1.5,0) [] {$\sfmod{}{\DiscMod{-2/3}^+}$};
\node (bot) at (0,-1.5) [] {$\IrrMod{0}$};
\node at (0,0) [nom] {$\StagMod{0}$};
\draw [->] (top) -- (left);
\draw [->] (top) -- (right);
\draw [->] (left) -- (bot);
\draw [->] (right) -- (bot);
\end{tikzpicture}
}.
\end{center}
Similarly, when $\lambda + \mu = \pm \tfrac{2}{3}$, we expect that $\TypMod{\pm 2/3}$ combines with $\sfmod{\mp 1}{\TypMod{\mp 2/3}}$ (again with appropriate superscripts) to form the indecomposable $\StagMod{\pm 2/3}^{\mp}$.  The other factor, $\sfmod{\pm 1}{\TypMod{0}}$, again splits off as a direct summand.  The relevant structure diagrams are therefore
\begin{center}
\parbox[c]{0.7\textwidth}{
\begin{tikzpicture}[thick,
	nom/.style={circle,draw=black!20,fill=black!20,inner sep=1pt}
	]
\node (top0) at (0,1.5) [] {$\DiscMod{2/3}^-$};
\node (left0) at (-1.5,0) [] {$\sfmod{-2}{\IrrMod{0}}$};
\node (right0) at (1.5,0) [] {$\sfmod{}{\IrrMod{0}}$};
\node (bot0) at (0,-1.5) [] {$\DiscMod{2/3}^-$};
\node (top1) at (6,1.5) [] {$\DiscMod{-2/3}^+$};
\node (left1) at (4.5,0) [] {$\sfmod{-1}{\IrrMod{0}}$};
\node (right1) at (7.5,0) [] {$\sfmod{2}{\IrrMod{0}}$};
\node (bot1) at (6,-1.5) [] {$\DiscMod{-2/3}^+$};
\node at (0,0) [nom] {$\StagMod{+2/3}^-$};
\node at (6,0) [nom] {$\StagMod{-2/3}^+$};
\draw [->] (top0) -- (left0);
\draw [->] (top0) -- (right0);
\draw [->] (left0) -- (bot0);
\draw [->] (right0) -- (bot0);
\draw [->] (top1) -- (left1);
\draw [->] (top1) -- (right1);
\draw [->] (left1) -- (bot1);
\draw [->] (right1) -- (bot1);
\end{tikzpicture}
}.
\end{center}
This confirms and refines the structural conjectures made in \cite{GabFus01}.\footnote{The structure of $\StagMod{-2/3}^+$ was also partially identified in \cite{AdaLat09} using an explicit construction based on lattice vertex algebras (similar to the free field methods of \cite{LesSU202,LesLog04}).  An intertwiner was also constructed there which is consistent with the second fusion rule of \eqref{FR:Gab}.}  Note that these diagrams become symmetric in terms of spectral flow indices upon regarding $\DiscMod{\mp 2/3}^{\pm}$ as having index $\pm \tfrac{1}{2}$.  We further remark (see \appref{app:Structure-4/3} for justifications) that these structure diagrams uniquely determine the $k=-\tfrac{4}{3}$ staggered modules, as \emph{admissible} modules, up to isomorphism.

The (conjectural) $\TypMod{}$-type fusion rules may therefore be summarised as follows:
\begin{equation} \label{FR:Typ-4/3}
\TypMod{\lambda} \fuse \TypMod{\mu} = 
\begin{cases}
\TypMod{0} \oplus \StagMod{0} & \text{if $\lambda + \mu = 0$,} \\
\sfmod{-1}{\TypMod{0}} \oplus \StagMod{-2/3}^+ & \text{if $\lambda + \mu = -\tfrac{2}{3}$,} \\
\sfmod{}{\TypMod{0}} \oplus \StagMod{2/3}^- & \text{if $\lambda + \mu = \tfrac{2}{3}$,} \\
\sfmod{-1}{\TypMod{\lambda + \mu - 4/3}} \oplus \TypMod{\lambda + \mu} \oplus \sfmod{}{\TypMod{\lambda + \mu + 4/3}} & \text{otherwise.}
\end{cases}
\end{equation}
Turning next to the (Grothendieck) fusion of the irreducible atypicals with typicals, we find that fusion with the vacuum and its images under spectral flow behaves as expected (as is easily verified).  We therefore turn to fusing with $\DiscMod{-2/3}^+$ and its spectrally-flowed versions.  This time, the Verlinde formula gives
\begin{equation}
\fuscoeff{\overline{\brac{\ell,-2/3}} \brac{m,\mu}}{\brac{n,\nu}} = \delta_{n = \ell + m} \func{\delta}{\nu = \mu - \tfrac{2}{3} \bmod{2}} + \delta_{n = \ell + m + 1} \func{\delta}{\nu = \mu + \tfrac{2}{3} \bmod{2}}.
\end{equation}
This yields, for example,
\begin{equation}
\ch{\DiscMod{-2/3}^+} \fuse \ch{\TypMod{0}} = \ch{\TypMod{-2/3}} + \ch{\sfmod{}{\TypMod{2/3}}} = \ch{\StagMod{-2/3}^+},
\end{equation}
which is consistent with \eqref{FR:Gab} and confirms the above structure diagrams.  However, it also yields
\begin{align}
\ch{\DiscMod{-2/3}^+} \fuse \ch{\StagMod{0}} &= \ch{\DiscMod{-2/3}^+} \fuse \Bigl( \ch{\sfmod{-1}{\TypMod{2/3}}} + \ch{\sfmod{}{\TypMod{-2/3}}} \Bigr) \notag \\
&= \ch{\sfmod{-1}{\TypMod{0}}} + \ch{\TypMod{-2/3}} + \ch{\sfmod{}{\TypMod{2/3}}} + \ch{\sfmod{2}{\TypMod{0}}} \notag \\
&= \ch{\sfmod{-1}{\TypMod{0}}} + \ch{\StagMod{-2/3}^+} + \ch{\sfmod{2}{\TypMod{0}}},
\end{align}
which is \emph{inconsistent} with the rule \eqref{FR:GabConj} conjectured in \cite{GabFus01}.  Indeed, this computation leads us to conjecture that the correct fusion rule is instead
\begin{equation} \label{FR:CRConj}
\DiscMod{-2/3}^+ \fuse \StagMod{0} = \sfmod{-1}{\TypMod{0}} \oplus \StagMod{-2/3}^+ \oplus \sfmod{2}{\TypMod{0}}.
\end{equation}
We remark that conformal dimensions do not rule out $\sfmod{-1}{\TypMod{0}}$ and $\sfmod{2}{\TypMod{0}}$ combining into an indecomposable.  However, we view this as unlikely.  Assuming \eqref{FR:CRConj}, associativity now implies that the fusion rules \eqref{FR:GabAss} must be replaced by
\begin{equation} \label{FR:CRAss}
%\begin{gathered}
%\TypMod{0} \fuse \StagMod{0} = \sfmod{-1}{\StagMod{2/3}^-} \oplus 2 \TypMod{0} \oplus \sfmod{}{\StagMod{-2/3}^+}, \qquad
%\TypMod{0} \fuse \StagMod{-2/3}^+ = \sfmod{-1}{\TypMod{0}} \oplus 2 \StagMod{-2/3}^+ \oplus \sfmod{2}{\TypMod{0}}, \\
%\DiscMod{2/3}^- \fuse \StagMod{-2/3}^+ = 2 \TypMod{0} \oplus \StagMod{0}, \qquad
%\StagMod{2/3}^- \fuse \StagMod{-2/3}^+ = \sfmod{-1}{\StagMod{2/3}^-} \oplus 4 \TypMod{0} \oplus 2 \StagMod{0} \oplus \sfmod{}{\StagMod{-2/3}^+}, \\
%\StagMod{0} \fuse \StagMod{0} = \sfmod{-3}{\TypMod{0}} \oplus \sfmod{-1}{\StagMod{2/3}^-} \oplus 2 \TypMod{0} \oplus 2 \StagMod{0} \oplus \sfmod{}{\StagMod{-2/3}^+} \oplus \sfmod{3}{\TypMod{0}}, \\
%\StagMod{0} \fuse \StagMod{-2/3}^+ = 2 \sfmod{-1}{\TypMod{0}} \oplus \sfmod{-1}{\StagMod{0}} \oplus 2 \StagMod{-2/3}^+ \oplus \sfmod{2}{\StagMod{0}} \oplus 2 \sfmod{2}{\TypMod{0}}.
%\end{gathered}
%\end{equation}
%\begin{equation}
\begin{split}
\DiscMod{2/3}^- \fuse \StagMod{-2/3}^+ &= 2 \TypMod{0} \oplus \StagMod{0}, \\
\TypMod{0} \fuse \StagMod{0} &= \sfmod{-1}{\StagMod{2/3}^-} \oplus 2 \TypMod{0} \oplus \sfmod{}{\StagMod{-2/3}^+}, \\
\TypMod{0} \fuse \StagMod{-2/3}^+ &= \sfmod{-1}{\TypMod{0}} \oplus 2 \StagMod{-2/3}^+ \oplus \sfmod{2}{\TypMod{0}}, \\
\StagMod{0} \fuse \StagMod{0} &= \sfmod{-3}{\TypMod{0}} \oplus \sfmod{-1}{\StagMod{2/3}^-} \oplus 2 \TypMod{0} \oplus 2 \StagMod{0} \oplus \sfmod{}{\StagMod{-2/3}^+} \oplus \sfmod{3}{\TypMod{0}}, \\
\StagMod{0} \fuse \StagMod{-2/3}^+ &= 2 \sfmod{-1}{\TypMod{0}} \oplus \sfmod{-1}{\StagMod{0}} \oplus 2 \StagMod{-2/3}^+ \oplus \sfmod{2}{\StagMod{0}} \oplus 2 \sfmod{2}{\TypMod{0}} \\
\StagMod{2/3}^- \fuse \StagMod{-2/3}^+ &= \sfmod{-1}{\StagMod{2/3}^-} \oplus 4 \TypMod{0} \oplus 2 \StagMod{0} \oplus \sfmod{}{\StagMod{-2/3}^+}.
\end{split}
\end{equation}
In computing these rules, we have first used the Verlinde formula to confirm that the fusion rule for $\DiscMod{2/3}^- \fuse \DiscMod{-2/3}^+$ given in \eqref{FR:Gab} is correct.

The conjectured fusion rule \eqref{FR:CRConj} nicely highlights the utility of the Verlinde formula.  Computing fusion rules directly using the Nahm-Gaberdiel-Kausch algorithm is very difficult in general.  A subtlety that deserves emphasising for such $\AKMA{sl}{2}$-computations is that one is often required to perform several such computations, determining different filtered quotients of the fusion product, in order to detect components of differing spectral flow index.  Worse yet, one has no \emph{a priori} knowledge concerning which of the infinitely many filtrations will lead to non-trivial contributions for the fusion product.  For example, the filtrations used in \cite{GabFus01} do not see the states of $\sfmod{-1}{\TypMod{0}}$ or $\sfmod{2}{\TypMod{0}}$, hence these modules were missed in the original conjecture \eqref{FR:GabConj}.  The Verlinde formula, however, yields this required \emph{a priori} knowledge effortlessly.  It predicts the spectral flow indices of the contributing modules and so tells us exactly which filtrations must be considered in order to deduce a complete picture of the fusion module and its indecomposable structure.

\subsection{Modular Invariants and Extended Algebras}

The search for bulk modular invariants is similar to the case of $k=-\tfrac{1}{2}$.  We again have the obvious diagonal and charge-conjugation partition functions, given by \eqref{MI:diag} and \eqref{MI:cc}, respectively, and their invariance again follows from the standard properties of the S-matrix.  There is also a family of discrete bulk modular invariant partition functions corresponding to simple current extensions of $\AKMA{sl}{2}_{-4/3}$.  The fusion rules reported in \eqref{FR:Gab}, \eqref{FR:Typ-4/3}, \eqref{FR:CRConj} and \eqref{FR:CRAss} show that the integer dimension simple currents again form a cyclic subgroup and that it is generated by $\sfmod{3}{\IrrMod{0}}$.

Taking therefore $b \in 3 \ZZ_+$, we may construct extended algebras $\WAlg{b}$, and $\TypMod{}$-type extended algebra modules $\WTypMod{\lambda}{b}$, which decompose into $\AKMA{sl}{2}_{-4/3}$-modules as follows:
\begin{equation}
\WAlg{b} \cong \sum_{\ell \in \ZZ} \sfmod{b \ell}{\IrrMod{0}}, \qquad 
\WTypMod{\lambda}{b} \cong \sum_{\ell \in \ZZ} \sfmod{b \ell}{\TypMod{\lambda}} \qquad \text{($b \in 3 \ZZ_+$).}
\end{equation}
The corresponding diagonal partition functions therefore take the form
\begin{equation}
\partfunc_b = \sum_{r=0}^{b-1} \sum_{s=0}^{b-1} \overline{\ch{\sfmod{r}{\WTypMod{2s/b}{b}}}} \ch{\sfmod{r}{\WTypMod{2s/b}{b}}} = \underset{\ell = m \bmod{b}}{\sum_{\ell \in \ZZ} \sum_{m \in \ZZ}} \sum_{j=0}^{b-1} \overline{\ch{\sfmod{\ell}{\TypMod{2j/b}}}} \ch{\sfmod{m}{\TypMod{2j/b}}}.
\end{equation}
T-invariance is easily checked and S-invariance may be demonstrated in an analogous fashion to the $k = -\tfrac{1}{2}$ case.  We therefore only repeat the main steps:  The modular S-transformation leads, as usual, to a sum over $\ell', m' \in \ZZ$ and an integral over $\lambda', \mu'$.  Performing the sums over $\ell$, $m=\ell + bn$ and $j$ leads to the factors
\begin{equation}
2 \func{\delta}{\lambda' = \tfrac{4}{3} \ell' + \mu' - \tfrac{4}{3} m' \bmod{2}}, \qquad \frac{2}{b} \func{\delta}{\mu' = \tfrac{4}{3} m' \bmod{\tfrac{2}{b}}}, \qquad b \delta_{\ell' = m' \bmod{b}},
\end{equation}
respectively, and simplifying recovers $\partfunc_b$.  We remark that $\partfunc_3$ is our candidate partition function for the $\AKMA{sl}{2}_{-4/3}$-theory whose $\AKMA{u}{1}$-coset is (related to) the $c=-7$ triplet model $W \brac{1,3}$.

%\begin{equation}
%\overline{\ch{\sfmod{\ell}{\TypMod{j/b}}}} \ch{\sfmod{m}{\TypMod{j/b}}}
%\end{equation}
%is T-invariant for all $j=0,...,2b-1$ if and only if
%\begin{equation}
%(m-\ell)(\frac{j}{b}-\frac{2}{3}(m+\ell)) =  0 \bmod{2}\, .
%\end{equation}
%We thus have two possibilities. Either $m=\ell \bmod 3b$ or $m=\ell\bmod{b}$ and $b$ is an integer multiple of $3$.
%Only the latter one turns out to lead to an S-invariant partition function, namely
%\begin{equation}
%\partfunc_b = \underset{\ell = m \bmod{b}}{\sum_{\ell \in \ZZ} \sum_{m \in \ZZ}} \sum_{j=0}^{b-1} \overline{\ch{\sfmod{\ell}{\TypMod{2j/b}}}} \ch{\sfmod{m}{\TypMod{2j/b}}}
%\end{equation}
%is modular invariant.  
%The computation is analoguous to the $k=-1/2$ case. For the reader's convenience we repeat the main steps.
%S-invariance follows immediately.

%\subsection{Extended Algebras???}

%The bulk modular invariant partition functions suggest the existence of extended algebras??
%Namely, we extend $\AKMA{sl}{2}_{-4/3}$ to an extended algebra $\mathcal A_b$, depending on above parameter $b$ as follows
%\begin{equation} 
%\mathcal A_b= \bigoplus_{n\in\mathbb Z}\sfmod{nb}{\IrrMod{0}}
%\end{equation}
%This algebra combines $\AKMA{sl}{2}_{-4/3}$ modules into extended algebra modules $M^\ell(\lambda)$ as follows.
%\begin{equation}
%M^\ell(\lambda) = \bigoplus_{n\in\mathbb Z} \sfmod{bn+\ell}{\TypMod{\lambda}}.
%\end{equation}
%Then the bulk modular invariant is a finite sum of extended algebra characters
%\begin{equation}
%\partfunc_b = \sum_{\ell =0}^{b-1} \sum_{j=0}^{b-1} \overline{\ch{M^\ell(2j/b)}} \ch{M^\ell(2j/b)}
%\end{equation}

\section*{Acknowledgements}

We would like to thank Simon Wood for valuable discussions relating to the results reported in this paper and for his detailed comments on a draft manuscript.  Thanks are also extended to Matthias Gaberdiel for correspondence concerning his $k=-\tfrac{4}{3}$ results.  In addition, we would like to thank the Institut Henri Poincar\'{e} and, in particular, the organisers of the recent trimestre ``Advanced Conformal Field Theory and Applications'', where some of this work was undertaken.  The research of DR is supported by an Australian Research Council Discovery Project DP0193910.

\appendix

\section{An Identity of Kac and Wakimoto} \label{app:Magic}

In this appendix, we will briefly review a remarkable identity of Kac and Wakimoto \cite[Eq.~(4.8)]{KacInt94} which turns out to be very useful for computing the characters of the indecomposable atypical $\AKMA{sl}{2}$-modules $\TypMod{\pm 1/2}^{\pm}$ in \secDref{sec:ModAtyp}{sec:Char-4/3}.  The identity in question is derived from the denominator formula for the affine Lie superalgebra $\AKMSA{sl}{2}{1}$:
\begin{multline}
\prod_{i=1}^{\infty} \frac{\brac{1-uvq^i} \brac{1-q^i}^2 \brac{1-u^{-1}v^{-1}q^{i-1}}}{\brac{1+uq^i} \brac{1+vq^i} \brac{1+u^{-1}q^{i-1}} \brac{1+v^{-1}q^{i-1}}} \\
= \frac{1-u^{-1}v^{-1}}{\brac{1+u^{-1}} \brac{1+v^{-1}}} - \sqbrac{\sum_{m,n = 1}^{\infty} - \sum_{m,n = -1}^{-\infty}} \brac{-1}^{m+n} u^m v^n q^{mn}.
\end{multline}
This equality holds for formal power series in $q$.  Both sides converge absolutely in the region where $\abs{q} < 1$ and $\abs{u} , \abs{v} < \abs{q}^{-1}$.  Kac and Wakimoto then interpret the right-hand side of this identity as a power series in $u$, obtaining
\begin{equation} \label{eqn:KW}
\prod_{i=1}^{\infty} \frac{\brac{1-uvq^i} \brac{1-q^i}^2 \brac{1-u^{-1}v^{-1}q^{i-1}}}{\brac{1+uq^i} \brac{1+vq^i} \brac{1+u^{-1}q^{i-1}} \brac{1+v^{-1}q^{i-1}}} = -\sum_{m \in \ZZ} \frac{\brac{-1}^m u^m}{1+vq^m}.
\end{equation}
The region of convergence is now $\abs{q} < \abs{u} < 1$.

We make the replacements $u \mapsto -u$ and $v \mapsto -v$ in \eqref{eqn:KW} in order to write the left-hand side more compactly.  The result is
\begin{equation} \label{eqn:Magic1}
\frac{\Jth{1}{uv ; q} \func{\eta}{q}^3}{\Jth{1}{u ; q} \Jth{1}{v ; q}} = -\ii \sum_{m \in \ZZ} \frac{u^m}{1-vq^m} \qquad \text{($\abs{q} < \abs{u} < 1$).}
\end{equation}
The region of convergence is important.  If we replace $u$ by $uq$ in the above, then we instead derive that
\begin{equation} \label{eqn:Magic2}
\frac{\Jth{1}{uv ; q} \func{\eta}{q}^3}{\Jth{1}{u ; q} \Jth{1}{v ; q}} = -\ii \sum_{m \in \ZZ} \frac{u^m vq^m}{1-vq^m} \qquad \text{($1 < \abs{u} < \abs{q}^{-1}$).}
\end{equation}
These identities will be used to expand characters, written as meromorphic functions, in the appropriate regions of convergence.  However, the results will be identified as expressing the characters as formal power series for which convergence issues may be neglected.

\section{The Structure of $k=-\tfrac{4}{3}$ Staggered Modules} \label{app:Structure-4/3}

In this appendix, we consider whether the structure diagrams deduced in \secref{sec:Verlinde-4/3} completely determine the staggered module up to isomorphism.  For $\StagMod{0}$, this may be settled using the method applied to its namesake at $k=-\tfrac{1}{2}$, as detailed in \cite[Sec.~4.3]{RidFus10}.  In brief, we choose $\ket{x_0^+}$ to span the weight space of $\StagMod{0}$ of weight $2$ and conformal dimension $-1$, then define $\ket{\omega_0} = f_{-1} \ket{x_0^+}$ and $\ket{x_0^-}$ by $e_{-1} \ket{x_0^-} = \ket{\omega_0}$.  Finally, the logarithmic partner $\ket{y_0}$ of $\ket{\omega_0}$ is normalised so that $L_0 \ket{y_0} = \ket{\omega_0}$.  The structure of $\StagMod{0}$ is then determined by the six constants appearing in the following equations:
\begin{equation}
\begin{aligned}
e_1 \ket{y_0} = \beta_0 \ket{x_0^+}, \\
f_1 \ket{y_0} = \tilde{\beta}_0 \ket{x_0^-},
\end{aligned}
\qquad
\begin{aligned}
e_0 \ket{y_0} = \beta'_0 h_{-1} \ket{x_0^+} + \beta''_0 f_{-2} e_1 \ket{x_0^+}, \\
f_0 \ket{y_0} = \tilde{\beta}'_0 h_{-1} \ket{x_0^-} + \tilde{\beta}''_0 e_{-2} f_1 \ket{x_0^-}.
\end{aligned}
\end{equation}
These constants are completely determined by the constraints $e_2 e_0 \ket{y_0} = f_2 f_0 \ket{y_0} = 0$, $h_1 e_0 \ket{y_0} = 2 e_1 \ket{y_0}$, $h_1 f_0 \ket{y_0} = -2 f_1 \ket{y_0}$, $e_0 f_0 \ket{y_0} = f_0 e_0 \ket{y_0}$ and $L_0 \ket{y_0} = \ket{\omega_0}$, yielding
\begin{equation}
\beta_0 = \tilde{\beta}_0 = \frac{1}{6}, \qquad -\beta'_0 = \tilde{\beta}'_0 = \frac{1}{2}, \qquad \beta''_0 = \tilde{\beta}''_0 = \frac{3}{4}.
\end{equation}
We conclude that $\StagMod{0}$ is determined up to isomorphism by its structure diagram.

That the staggered module $\StagMod{0}$ is admissible, meaning that it is a module of the vertex operator algebra, can be checked by acting on $\ket{y_0}$ with the zero-mode of the null field
\begin{equation}
\chi = -36 \normord{ehf} + 24 \normord{e \partial f} + 9 \normord{hhh} - 96 \normord{\partial e f} - 18 \normord{\partial h h} + 4 \partial^2 h,
\end{equation}
which corresponds to the $f_0$-descendant of the (non-trivial) singular vector of the vacuum Verma module.  The result is $\chi_0 \ket{y_0} = 48 \tbrac{\beta_0 - \tilde{\beta}_0} \ket{\omega_0} = 0$ as required.  We remark that the conclusion $\beta_0 = \tilde{\beta}_0$ has a curious consequence:  From
\begin{equation}
\beta_0 \braket{x_0^+}{x_0^+} = \braket{\omega_0}{y_0} = \tilde{\beta}_0 \braket{x_0^-}{x_0^-},
\end{equation}
we conclude that the norms of $\ket{x_0^+}$ and $\ket{x_0^-}$ must be taken as equal.  In other words, we are not free to choose the relative normalisation of these vectors, as one might have na\"{\i}vely supposed.

The analysis for $\StagMod{-2/3}^+$ is somewhat more subtle.  Now, we choose $\ket{x_{-2/3}^+}$ to span the weight space of weight $\tfrac{4}{3}$ and conformal dimension $-\tfrac{1}{3}$, and then define $\ket{\omega_{-2/3}} = f_0 \ket{x_{-2/3}^+}$ and $\ket{x_{-2/3}^-}$ by $e_{-1} \ket{x_{-2/3}^-} = \ket{\omega_{-2/3}}$.  The logarithmic partner to $\ket{\omega_{-2/3}}$ will be denoted by $\ket{y_{-2/3}}$ and is normalised as usual so that $\brac{L_0 + \tfrac{1}{3}} \ket{y_{-2/3}} = \ket{\omega_{-2/3}}$.  We therefore have only two constants to consider, defined by
\begin{equation}
e_0 \ket{y_{-2/3}} = \beta_{-2/3} \ket{x_{-2/3}^+}, \qquad f_1 \ket{y_{-2/3}} = \tilde{\beta}_{-2/3} \ket{x_{-2/3}^-}.
\end{equation}
The only obvious constraint, however, is $\brac{L_0 + \tfrac{1}{3}} \ket{y_{-2/3}} = \ket{\omega_{-2/3}}$ which yields $\beta_{-2/3} + \tilde{\beta}_{-2/3} = -\tfrac{2}{3}$.  It therefore appears that there may exist a one-parameter family of $\AKMA{sl}{2}_{-4/3}$-modules sharing the structure diagram of $\StagMod{-2/3}^+$.\footnote{We will not prove that this is the case (or not).  We expect that this can be settled by using a variant of the singular vector arguments introduced in \cite{RidLog07,RidSta09}.}  But, we have not yet imposed the requirement of admissibility.  Doing so, one obtains $\chi_0 \ket{y_{-2/3}} = 24 \tbrac{\beta_{-2/3} - \tilde{\beta}_{-2/3}} \ket{\omega_0} = 0$, hence that
\begin{equation}
\beta_{-2/3} = \tilde{\beta}_{-2/3} = -\tfrac{1}{3}.
\end{equation}
We conclude that $\StagMod{-2/3}$ is determined, as an admissible $\AKMA{sl}{2}_{-4/3}$-module, up to isomorphism by its structure diagram.  Note that this again forces the norms of $\ket{x_{-2/3}^+}$ and $\ket{x_{-2/3}^-}$ to coincide.

\raggedright

%\bibliography{mod}
%\bibliographystyle{unsrt}

\end{document}